\documentclass[prb,twocolumn,superscriptaddress,showpacs]{revtex4-1}
\usepackage{amsfonts}
\usepackage{amsmath}
\usepackage{amssymb}
\usepackage{bm}
\usepackage{color}
\usepackage[usenames,dvipsnames,svgnames,table]{xcolor}
\usepackage{graphicx}
\usepackage[varg]{txfonts}
\usepackage{wrapfig}
\usepackage{varwidth}
\usepackage{dcolumn}
\usepackage{mathrsfs}
\usepackage[breaklinks,colorlinks = true,linkcolor = blue,urlcolor=cyan,citecolor=blue]{hyperref}
\usepackage{soul}

\usepackage{times}
\usepackage{subfigure}
\begin{document}

\title{Acoustic signatures of the phases and phase transitions in  Yb$_2$Ti$_2$O$_7$}

\author{Subhro ~Bhattacharjee} 
\affiliation{Max-Planck-Institut f\"ur Physik komplexer Systeme, D-01187 Dresden, Germany} 
\affiliation{International Centre for Theoretical Sciences, Tata Institute of Fundamental Research, Bangalore 560012, India}

\author{S.~Erfanifam} 
\affiliation{Hochfeld-Magnetlabor Dresden (HLD-EMFL), Helmholtz-Zentrum Dresden-Rossendorf, D-01314 Dresden, Germany}

\author{E.L.~Green}
\affiliation{Hochfeld-Magnetlabor Dresden (HLD-EMFL), Helmholtz-Zentrum Dresden-Rossendorf, D-01314 Dresden, Germany}

\author{M.~Naumann}
\affiliation{Hochfeld-Magnetlabor Dresden (HLD-EMFL), Helmholtz-Zentrum Dresden-Rossendorf, D-01314 Dresden, Germany}
\affiliation{Institut f\"ur Festk\"orperphysik, Technische Universit\"at Dresden, D-01062 Dresden, Germany}

\author{Zhaosheng Wang}
\affiliation{Hochfeld-Magnetlabor Dresden (HLD-EMFL), Helmholtz-Zentrum Dresden-Rossendorf, D-01314 Dresden, Germany}

\author{S.~Granovski}
\affiliation{Institut f\"ur Festk\"orperphysik, Technische Universit\"at Dresden, D-01062 Dresden, Germany}

\author{M.~Doerr}
\affiliation{Institut f\"ur Festk\"orperphysik, Technische Universit\"at Dresden, D-01062 Dresden, Germany}

\author{J.~Wosnitza} 
\affiliation{Hochfeld-Magnetlabor Dresden (HLD-EMFL), Helmholtz-Zentrum Dresden-Rossendorf, D-01314 Dresden, Germany}
\affiliation{Institut f\"ur Festk\"orperphysik, Technische Universit\"at Dresden, D-01062 Dresden, Germany}

\author{A.A.~Zvyagin} 
\affiliation{Max-Planck-Institut f\"ur Physik komplexer Systeme, D-01187 Dresden, Germany}
\affiliation{B.I.~Verkin Institute for Low Temperature Physics and Engineering of the National Academy of Sciences of Ukraine, Kharkov, 61103, Ukraine}

\author{R.~Moessner} 
\affiliation{Max-Planck-Institut f\"ur Physik komplexer Systeme, D-01187 Dresden, Germany} 

\author{A.~Maljuk} 
\affiliation{Leibniz Institute for Solid State and Materials Research Dresden, 01069 Dresden, Germany}

\author{S.~Wurmehl} 
\affiliation{Institut f\"ur Festk\"orperphysik, Technische Universit\"at Dresden, D-01062 Dresden, Germany}
\affiliation{Leibniz Institute for Solid State and Materials Research Dresden, 01069 Dresden, Germany}

\author{B.~B\"uchner}
\affiliation{Institut f\"ur Festk\"orperphysik, Technische Universit\"at Dresden, D-01062 Dresden, Germany} 
\affiliation{Leibniz Institute for Solid State and Materials Research Dresden, 01069 Dresden, Germany}

\author{S.~Zherlitsyn} 
\affiliation{Hochfeld-Magnetlabor Dresden (HLD-EMFL), Helmholtz-Zentrum Dresden-Rossendorf, D-01314 Dresden, Germany}

\begin{abstract}
We report on measurements of the sound velocity and attenuation in a single crystal of the candidate quantum-spin-ice material Yb$_2$Ti$_2$O$_7$ as a function of temperature and magnetic field. The acoustic modes couple to the spins magneto-elastically and, hence, carry information about the spin correlations that sheds light on the intricate  magnetic phase diagram of Yb$_2$Ti$_2$O$_7$ and the nature of spin dynamics in the material. Particularly, we find a pronounced thermal hysteresis in the acoustic data with a concomitant peak in the specific heat indicating a possible first-order phase transition at about 0.17 K. At low temperatures, the acoustic response to magnetic field saturates hinting at the development of magnetic order. Furthermore, mean-field calculations suggest that Yb$_2$Ti$_2$O$_7$  undergoes a first-order phase transition from a cooperative paramagnetic phase to a ferromagnet below $T \approx 0.17$ K.
\end{abstract}

\pacs{75.50.-y, 72.55.+s}
\date{\today}
\maketitle

\section{Introduction}

It has been proposed that a class of rare-earth pyrochlore magnets, dubbed quantum spin ice, can host Coulombic quantum spin-liquid\cite{hermele2004pyrochlore,PhysRevB.68.054405} phases at low temperatures. Such phases support novel emergent excitations, such as photons and magnetic monopoles. \cite{gingras2014quantum,hermele2004pyrochlore,savary2012coulombic,castelnovo2008magnetic,PhysRevLett.108.067204,PhysRevB.86.075154,PhysRevB.68.054405,PhysRevLett.91.167004,henley2010coulomb}  This has led to an active search for material examples exhibiting ``emergent quantum electrodynamics" in condensed-matter systems.\cite{chang2012higgs,gingras2014quantum,applegate2012vindication,thompson2011rods,ross2011quantum,yin2013low,fennell2012power,legl2012vibrating,hermele2004pyrochlore,savary2012coulombic,castelnovo2008magnetic,hermele2004pyrochlore,savary2012coulombic,lee2012generic,pan2015measure,PhysRevLett.80.2933,PhysRevLett.108.067204,PhysRevB.86.075154,PhysRevLett.108.247210} An essential ingredient for the proposed rich phenomena is quantum-mechanical tunnelling between the macroscopically degenerate classic spin-ice (``2-in-2-out") ground-state spin configurations.

In this context, Yb$_{2}$Ti$_{2}$O$_{7}$ has emerged as one of the leading candidates for a quantum-spin-ice realization. In Yb$_{2}$Ti$_{2}$O$_{7}$, the crystal field surrounding the Yb$^{+3}$ ions, that form the pyrochlore spin-network, stabilizes a well-separated $S=1/2$ Kramers doublet which accounts for the low-energy magnetic properties.\cite{hodges2001crystal} Recent experiments reveal that in addition to the dominant frustrated (classic) Ising exchange interactions along the local [111] directions, significant transverse spin exchanges are present in Yb$_{2}$Ti$_{2}$O$_{7}$.\cite{chang2012higgs,applegate2012vindication,thompson2011rods,ross2011quantum} The exact role of the transverse terms in shaping the low-energy properties of Yb$_{2}$Ti$_{2}$O$_{7}$ is, however, far from understood. In the temperature window between approximately 0.24 and 1 K, recent experiments indicate that Yb$_{2}$Ti$_{2}$O$_{7}$ is in a correlated paramagnetic phase.\cite{chang2012higgs,thompson2011rods,ross2011quantum} However, whether this phase extends down to lowest temperatures giving rise to a quantum-spin-ice ground state is a topic of active current research. 

Notably, aside from the mentioned Coulomb spin liquid, the transverse terms can potentially lead to magnetic ordering as they lift the classical degeneracy.\cite{chang2012higgs,yasui2003ferromagnetic} Indeed, recent muon-spin relaxation,\cite{chang2014static} magnetization,\cite{lhotel2014first} magnetic susceptibility,\cite{dun2014chemical} and single-crystal neutron-scattering\cite{yasui2003ferromagnetic} measurements show evidence of long-range ferromagnetic spin correlations below $T\approx 0.24$ K with a reduced ordered moment.  However, a host of other experiments have failed to detect such low-temperature ferromagnetic order,\cite{PhysRevB.84.174442,PhysRevLett.88.077204,PhysRevB.70.180404,PhysRevLett.103.227202,PhysRevB.88.134428} although some short-range correlations have been observed and the paramagnetic state was also found to transform into a fully polarized state on applying a modest magnetic field.\cite{PhysRevLett.103.227202} Furthermore, the low-temperature properties of Yb$_{2}$Ti$_{2}$O$_{7}$ are known to be sensitive to small variations in stoichiometry depending on the oxygen content and possible Yb$^{3+}$ substitutions at  Ti$^{4+}$ positions (``stuffing"). \cite{PhysRevB.86.174424,chang2012higgs,PhysRevB.84.172408} From the theoretical point of view several exotic scenarios have been suggested for the phase below $T\approx 0.24$ K including a first-order transition from a quantum-spin-ice state to a ferromagnet associated with the condensation of magnetic monopoles \cite{chang2012higgs} or a low-temperature Coulomb ferromagnetic phase that has finite magnetization as well as gapless emergent photons.\cite{savary2012coulombic,PhysRevB.69.224415} 

In this work, we were able to obtain a complementary insight into the low-temperature spin physics of Yb$_{2}$Ti$_{2}$O$_{7}$ by probing the elastic modes which couple to the spins through magneto-elastic coupling and, hence, can reveal information regarding the nature of the magnetic state, particularly the spin fluctuations.  Indeed, previous studies showed that the coupling between the phonons and spins in classical spin ice could lead to distinct signatures in ultrasonic measurements which shed light on the low-temperature state and particularly its excitations (both in the presence and absence of an external magnetic field).\cite{PhysRevB.84.220404,PhysRevB.90.064409} Such investigations, however, have so far rarely been done for quantum-spin-ice candidates. Here, we report on pronounced features in the acoustic properties of Yb$_{2}$Ti$_{2}$O$_{7}$ at low temperatures that arise due to  magneto-elastic coupling. In particular, our results show evidence for spontaneous and field-induced phase transformations. In zero magnetic field, we find a softening of the sound velocity (and a simultaneous rise in the sound attenuation) with decreasing temperature due to the development of spin-spin correlations in the compound. The qualitative character of the softening changes at around 0.17 K and shows pronounced thermal hysteresis (Fig. \ref{YbTOT dependcT}). At low temperatures, the sound velocity (attenuation) shows non-monotonic behaviour as a function of magnetic field and a clear minimum (maximum) is observed at and above $T\sim 0.17$ K (Figs. \ref{YbTOB dependcT} and \ref{YbTOB dependcL}). For lower temperatures, the behaviour is monotonic. For all cases the velocity and attenuation saturate for fields greater than about $2.2$ T indicating complete polarization of the magnetic moments. Using a mean-field calculation, our results suggest that Yb$_{2}$Ti$_{2}$O$_{7}$ undergoes a first-order transition into a ferromagnetic state from a cooperative paramagnet at $T\approx 0.17$ K, where a peak in the specific heat is also observed in conformity with earlier results.\cite{chang2012higgs,yasui2003ferromagnetic} The thermal hysteresis observed in the sound velocity and attenuation is a direct result of this first order transition. Using these insights, we draw an empirical phase diagram for Yb$_{2}$Ti$_{2}$O$_{7}$ as a function of temperature and magnetic field.


\section{Experimental details}

\subsection{Sample preparation}
Stoichiometric quantities of Yb$_{2}$O$_{3}$ (4N) and TiO$_{2}$ (4N) were well mixed in ethanol, dried and pre-synthesized at 1000~$^{\circ}\mathrm{C}$ for 24~h in air. Commercially available Yb$_{2}$O$_{3}$ was pre-heated at 900~$^{\circ}\mathrm{C}$ for 10 h in air to get rid of water. The pre-sintered powder was pressed into rods under 3 kbar pressure using a cold isostatic press (EPSI) and finally synthesized at 1400~$^{\circ}\mathrm{C}$ for 24 h in oxygen flow. The crystal was grown at 5 mm/h and 4 bar oxygen pressure by a crucible-free floating-zone method with optical heating using a 4-mirror-type image furnace. The crystals had a typical diameter of 5.5 - 6.0 mm and a length up to 20 mm, with yellowish color and well transparent in the visible light. The investigated Yb$_{2}$Ti$_{2}$O$_{7}$ single crystals have a cubic crystal structure with space group $Fd\overline{3}m$ consistent with literature \cite{PhysRevB.86.174424} as confirmed by powder and single-crystal diffractometry. The microstructure of the crystal was characterized by scanning electron microscopy (Zeiss EVOMA15) along with compositional analysis using an electron microprobe analyzer for semi-quantitative elemental analysis in the energy-dispersive x-ray (EDX) mode (X-MaxN20 detector from Oxford Instruments with AZtecEnergy Advanced acquisition and EDX analysis software). The composition is determined by averaging over several points (on the order of 10 points). The crystal was found homogeneous in composition and had no visible inclusions as confirmed by EDX analysis and optical microscopy. The composition of the crystal is stoichiometric with respect to Yb and Ti within the typical error bars of an EDX experiment. The systematic error typically given for EDX analysis is on the order of 1-2 ~\%. It is known that the long-range ferromagnetic order in Yb$_{2}$Ti$_{2}$O$_{7}$ is quite sensitive to Yb deficiency.\cite{chang2012higgs} Please note that the measurement of the oxygen content is not reliable with EDX. 

\subsection{Details of the measurements}

Two parallel (111) surfaces were cut and polished giving a sample length of about 3.7~mm for the ultrasound-propagation experiments with a size of a few millimeters in the plane normal to the sound-propagation direction. The sample orientation was checked using x-ray Laue technique. Two acoustic modes have been studied, the longitudinal $c_L$, $c_{L} = (c_{11}+2c_{12}+4c_{44})/3$ with {\bf k}$\|${\bf u}$\|$[111] and the transverse $c_T$ mode, $c_T = (c_{11}+c_{44}-c_{12})/3$ ({\bf k}$\|$[111], {\bf u}$\bot${\bf k}). Here, {\bf k} and {\bf u} are the wave vector and polarization of the acoustic wave, respectively. The elastic constants $c_{\Gamma}$ are related to the sound velocity, $v_{\Gamma}$, $c_{\Gamma}=\rho v_{\Gamma}^{2}$, where $\rho$ is the mass density. Resonance LiNbO$_{3}$ and wide-band PVDF-film (polyvinylidene fluoride film) transducers glued to the sample were employed for sound generation and detection. The relative change of the sound velocity, $\Delta v/v$, and the sound attenuation, $\Delta \alpha$, were obtained using a phase-sensitive-detection technique.\cite{luthi2007physical}  Calibrated RuO$_{2}$ resistor, directly attached to the sample was employed for thermometery. We used a dilution refrigerator (0.02 - 2.5~K) combined with a commercial 20~Tesla superconducting magnet. Before each field sweep, we demagnetized the magnet at sample temperature of $\approx 2.5$~K and then cooled the sample to the desired temperature. The measurements were performed under the zero-field-cooled (ZFC) conditions.

Thermal expansion of Yb$_2$Ti$_2$O$_7$ was studied along the [111] direction by capacitive dilatometry in the temperature range between 60~mK and 1.5~K using a tilted-plate miniaturized dilatometer~\cite{rotter}. To avoid thermal gradients the dilatometer was positioned directly into the mixing chamber of a top-loading dilution refrigerator. Vibrations were reduced by damping the pumping lines. All these arrangements result in a resolution of relative length changes of $10^{-9}$. 

Specific heat was measured using the heat-pulse method. Measurements were carried out in a 3He/4He top-loading dilution refrigerator in combination with a 16/18 T superconducting magnet. The magnet was first warmed to temperatures above 100 K to ensure an accurate zero field measurement. A 332.7 $\mu$g piece of Yb$_2$Ti$_2$O$_7$, from the same sample used for ultrasound measurements, was secured on a sapphire plate with Apiezon N-grease. The addenda, including the contribution from the N-grease and the sapphire plate, was accurately measured and subtracted from the obtained results. A Lake Shore 370 AC resistance bridge was used to monitor a calibrated thin film RuO$_2$ temperature sensor and a Yokogawa DC current source was used to apply a heat pulse (current ranged from 1-25 $\mu$A for
100-300 ms) to the 2 kOhm heater. Both the temperature sensor and the heater were fixed to the bottom of the sapphire plate which was suspended by nylon threads in order to provide a small heat link to the cold bath.

\section{Results and Discussion}

\subsection{Temperature dependences}

\begin{figure}
\begin{center}
\includegraphics[scale=0.41]{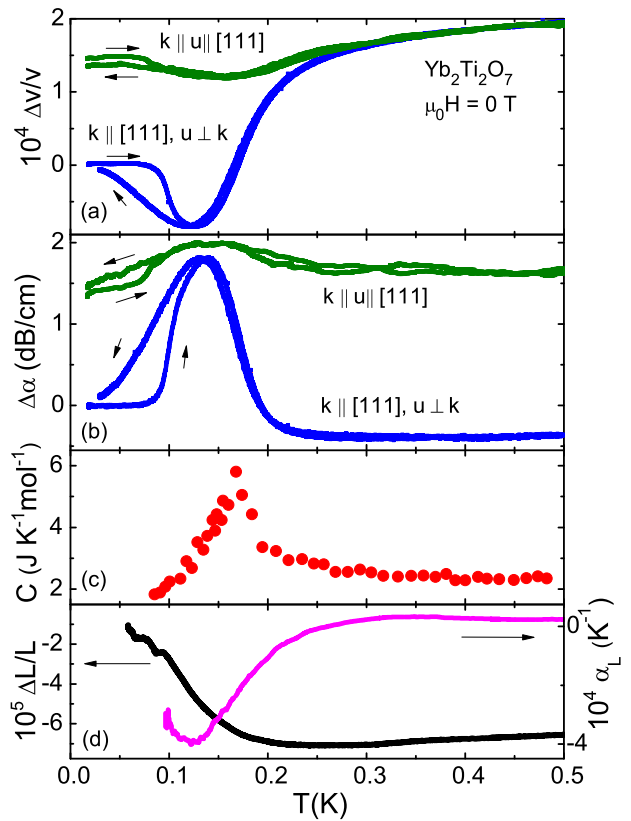}
\end{center}
\caption{(Color online) Temperature dependence of (a) sound velocity, $\Delta v/v$, and (b) sound attenuation, $\Delta \alpha$, for transverse $c_{T}$ and longitudinal $c_{L}$ acoustic modes, propagating along the [111] direction. Ultrasound frequency is  94.4 and 110.8 MHz for the longitudinal and transverse mode respectively. (c) Temperature dependence of the specific heat. (d) Temperature dependence of the thermal expansion, $\Delta L/L$, and the linear thermal expansion  coefficient, $\alpha_{L}$, along the [111] direction.}
\label{YbTOT dependcT}
\end{figure}

In Fig.~\ref{YbTOT dependcT}(a), we show the sound-velocity change ($\Delta v/v$) as a function of temperature (below $0.5$ K) for both the longitudinal and transverse polarizations.  Upon cooling $\Delta v/v$ decreases untill it reaches a minimum at around 130~mK. Concomitantly, the sound attenuation ($\Delta\alpha$) increases rapidly below 220 mK revealing a maximum at about 130~mK [Fig.~\ref{YbTOT dependcT}(b)]. Below this temperature, both $\Delta v/v$ and $\Delta\alpha$ show pronounced hysterises down to the lowest measured temperatures.
In the longitudinal acoustic mode the effects are much smaller than for the transverse mode. This may occur due to (i) indirect exchange interactions being dominant in Yb$_{2}$Ti$_{2}$O$_{7}$, and, (ii) the smaller velocity of the transverse mode [which occurs in the denominator of the expression for $\Delta v/v$, see Eq. (15) in Ref. \onlinecite{PhysRevB.83.184421}].

In Fig.~\ref{YbTOT dependcT}(c), we plot the temperature dependence of the specific heat of the single crystal where we see a clear anomaly at about $170$ mK. This result is  in agreement with earlier specific-heat data obtained for Yb$_{2}$Ti$_{2}$O$_{7}$ single crystals as reported by Chang et al.\cite{chang2012higgs} The onset of the specific-heat anomaly coincides with the rapid changes in the sound attenuation and velocity around 220 mK, whereas the maximum  in the specific heat correlates with the inflection points for both, the sound velocity and attenuation. Most notably, the hysteresis appears at temperatures lower than the specific-heat peak. 

Finally, the thermal expansion, $\Delta L/L$, [Fig.~\ref{YbTOT dependcT}(d)] exhibits a strong increase upon cooling below 200 mK without saturation down to lowest measured temperatures. The linear thermal expansion coefficient, $\alpha_{L}$, which is giving by the temperature derivative of the thermal expansion, resembles closely the sound-velocity temperature dependence with a minimum at about 130~mK. Note, that $\Delta L/L$ is smaller than the sound-velocity change at these temperatures. Hence, the sound-velocity anomalies cannot be explained by the thermal expansion alone. 

In an insulating magnet, changes in the sound velocity and attenuation at low temperatures typically occur due to the coupling of the phonons to spin fluctuations. This is caused (i) since the sound waves modulate the position of the magnetic and non-magnetic ions (involved in the super-exchange) and, hence, modify the effective magnetic interactions, and (ii) by alterations of the local crystal field by changing the position of the ligands (nonmagnetic ions surrounding the magnetic one). This affects the single-ion properties of the magnetic ion (e.g., the effective $g$-factor and the single-ion magnetic anisotropy).
For weak sound waves and low temperatures and small magnetic fields (pertinent to our case), the first mechanism is dominant.\cite{Tachiki01011974} The changes in the interactions between magnetic ions, in turn, renormalize the sound velocity and attenuation. These renormalizations are then proportional to various spin-correlation functions\cite{PhysRevB.83.184421} (see below). Hence, measuring the sound velocity and attenuation, both as a function of temperature and magnetic field, can lead to a wealth of information about the nature of the magnetic correlations present in the system.

The observed features, as we shall show below, owe their origin to changes in the spin correlations as a function of temperature. To characterize the experimental observations, we generalize the calculations of Ref. \onlinecite{PhysRevB.83.184421} where it was shown that for cubic pyrochlores,
\begin{equation}
{\Delta v\over v} \approx A (\varepsilon_{spin} -DTc_{spin}) \ , 
\label{eq1}
\end{equation} 
where both the magneto-elastic parameters $A$ and $D$ depend on the details of polarization and direction of the sound-wave propagation as well as on the sound velocity, and derivatives of the exchange integrals with respect to the change of the distances between magnetic ions. $\varepsilon_{spin}$ and $c_{spin}$ are the average energy per spin and the spin contribution to the specific heat, respectively (See Appendix \ref{appen_spen}). 

Hence, as mentioned above, the acoustic signatures depend on the spin-correlation functions, namely two-spin and four-spin correlators. Approximate calculations of these correlators suitable for different phases then can be compared to the present experiments in order to obtain a better understanding of the magnetic phase. 

Further, following Ref. \onlinecite{Tachiki01011974}, (only keeping the ${\bf q }$ = 0 component of the magnetic susceptibility, $\chi$), the sound attenuation can be related to the sound-velocity change as
\begin{equation}
\Delta \alpha \approx -{\delta v\over v} {\omega_k^2\over v}{2BT\chi\over 4B^2 T^2\chi^2+\omega_k^2} \ , 
\label{eq2}
\end{equation}
where $\omega_k(=v k)$ is phonon frequency ($v$ is the velocity of sound and $k$ is the momentum) and $B$ is a material-dependent constant, respectively. 

\begin{figure}
\centering
\includegraphics[scale=0.41]{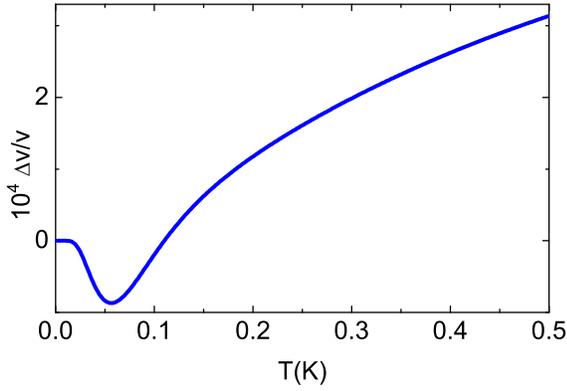}
\caption{(Color online) Sound-velocity change as obtained for classical spin ice within single tetrahedron approximation. The sound attenuation is negative of the fractional change in sound velocity (see text). The parameters used [with reference to Eq.~(\ref{eq1}), up to an arbitrary shift] are $A=1, D=1,J=0.07$.}
\label{fig_theory1}
\end{figure}

{We find that the qualitative features observed in our experiments on Yb$_{2}$Ti$_{2}$O$_{7}$, except for the hysteresis, can be reproduced within the single-tetrahedron approximation of a classical spin ice. The expressions for the specific heat and energy per spin are given in Appendix  \ref{appen_spen}. Using these expressions and approximating the magnetic susceptibility by its homogeneous value, $\chi$, for classical spin ice as that of a gas of monopoles \cite{ryzhkin2005magnetic,bramwell2012generalized} (see also the review article Ref.~\onlinecite{zvyagin2013new}) giving 
\begin{equation}
\chi = {\sqrt{3} \mu_0^2 \over a^3 k_B T} \ , 
\label{eq3}
\end{equation}
(where $a$ is the lattice constant) we find using Eq. (\ref{eq2}), $\Delta\alpha\propto -\Delta v/v$, {\it i.e.}, the sound attenuation in spin ice is proportional to the sound-velocity change as indeed observed experimentally. The temperature dependence of $\Delta v/v$  within our approximation is plotted in Fig. \ref{fig_theory1}.}

Our model can describe the qualitative behavior of the sound velocity, but deviates from the experiment quantitatively. This is in accordance with the fact that we have not taken into account the transverse terms which are substantial in the actual material. Further, as noted earlier, the presence of hysteresis, observed in experiments below $T\sim 0.2$ K, is also not explained within this approximation. Indeed, hystereses have as well been observed in magnetization data in Yb$_2$Ti$_2$O$_7$. \cite{lhotel2014first} Taking clue from these magnetization results along with neutron-scattering measurements, that show enhanced ferromagnetic correlations below $T\sim 0.2$ K,\cite{chang2012higgs,yasui2003ferromagnetic} we  argue that Yb$_2$Ti$_2$O$_7$ undergoes a first-order transition from a cooperative paramagnet to a ferromagnet. Notably, previous muon- and M\"{o}ssbauer-spectroscopy measurements suggest that around the same temperature there is a sudden suppression of spin fluctuation rate from $~10^3$ MHz above $0.2$ K to $1$ MHz below it.\cite{yaouanc2003spin,chang2012higgs,yasui2003ferromagnetic} In the present experiments the sound modes have a frequency of about 100 MHz. Thus, the sound modes essentially see a frozen spin background and only the first term in Eq. (\ref{eq1}) contributes to the fractional change in the sound velocity. 

\subsection{Sound velocity and attenuation vs. magnetic field}

\begin{figure}
\begin{center}
\includegraphics[scale=0.41]{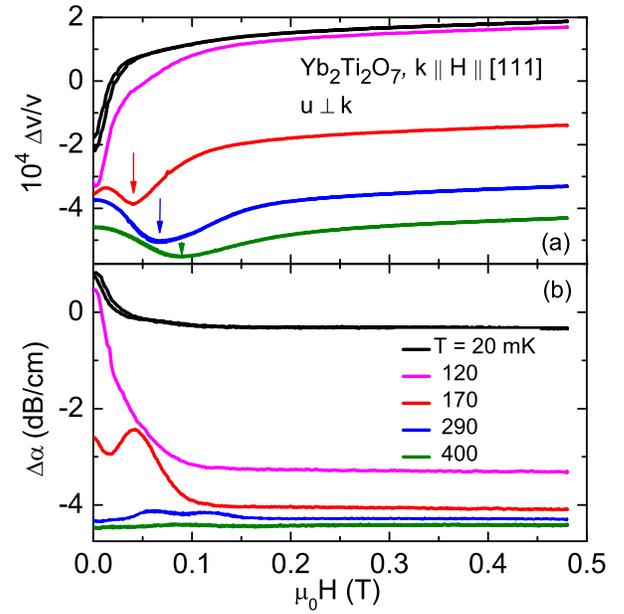}
\end{center}
\caption{(Color online) Field dependence of (a) the sound velocity, $\Delta v/v$, and (b) the sound attenuation, $\Delta \alpha$, for the transverse acoustic mode $c_{T}$, propagating along the [111] direction. The magnetic field is applied along the same direction. Arrows indicate the minimum positions in the sound velocity. Except for 0.02~K, data are shown for increasing field. The curves are arbitrarily shifted along the $y$ axis for clarity. The used ultrasound frequency was 110.8~MHz.}
\label{YbTOB dependcT}
\end{figure}

\begin{figure}
\begin{center}
\includegraphics[scale=0.41]{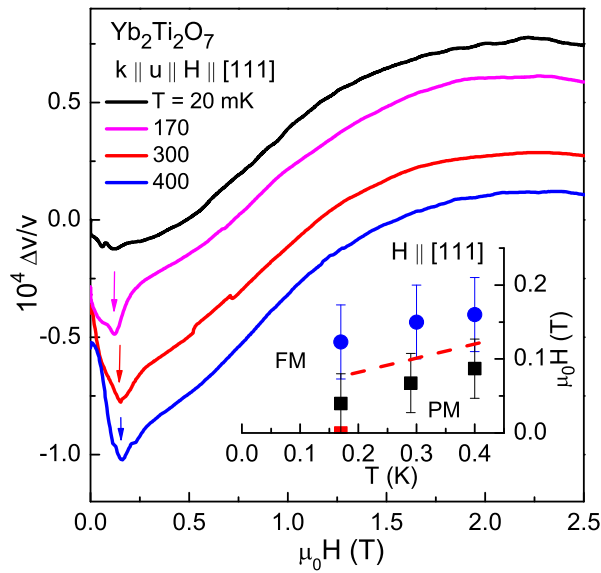}
\end{center}
\caption{(Color online) Field dependence of the sound velocity, $\Delta v/v$, for the longitudinal acoustic mode $c_{L}$, propagating along the [111] direction in Yb$_{2}$Ti$_{2}$O$_{7}$. Magnetic field is applied along the same direction. The arrows mark position of anomalies in the low-field region. The inset presents the low-temperature, low-field part of the $H-T$ phase diagram extracted from our ultrasound and specific-heat measurements for magnetic field applied along [111] direction. Here FM and PM, mean ferromagnet and paramagnet. See text for details.}
\label{YbTOB dependcL}
\end{figure}

Figure~\ref{YbTOB dependcT} shows a field dependence of the sound velocity and attenuation at different temperatures for the transverse sound mode.  At the two lowest temperatures, 20 and 120~mK, the sound velocity increases accompanied by a decrease in the sound attenuation. At 170 mK and above, a minimum in the sound velocity appears with a corresponding maximum in the attenuation. Interestingly, the changes in the acoustic properties at 0.5 T are somewhat larger than the temperature-induced variations (Fig.~\ref{YbTOT dependcT}). This can be attributed to the fact that the magnetic field suppresses quantum fluctuations, counterbalancing their effects. We also note that the hysteresis seen in the field sweeps is very small (only prominent in the 20 mK data). 

\begin{figure}
\centering
\includegraphics[scale=0.41]{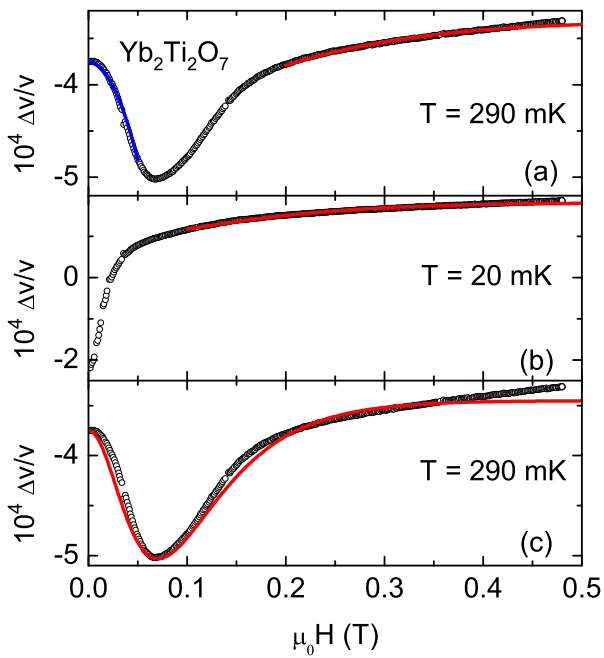}
\caption{(Color online) (a) and (b) data and fits of the transverse-mode $\Delta v/v$ versus magnetic field using the asymptotic functions of Eqs. (\ref{low}) (blue line for the paramagnetic phase) and (\ref{high}) (red lines for the polarized phase). (c) red fit curve using a mean-field theory (see text) describing the experimental data at 290 mK. The fitting parameters are discussed in Appendix \ref{appen_fit}.}
\label{fit}
\end{figure}

Figure~\ref{YbTOB dependcL} shows the field dependence of the longitudinal sound velocity at various temperatures. Again, similar to the transverse mode, a minimum (marked by arrows) is seen below 0.2~T for temperatures close to or above the specific-heat anomaly. However, the low-field anomalies become sharper at higher temperatures, although the corresponding anomalies broaden for the transverse mode (Fig.~\ref{YbTOB dependcT}).

Above the phase transition, in the paramagnetic phase, we can derive the field dependence of the sound velocity for small magnetic fields from Eq.~(\ref{eq1}) by expanding the spin correlations in powers of the field (see Appendix \ref{appen_derive}). We obtain
\begin{align}
\frac{\Delta v}{v}\sim{\rm constant}+A''h^2 ,
\label{low}
\end{align}
where $A''$ depends on the magnetic susceptibility and temperature. Similarly, for high fields (Appendix \ref{appen_derive}), when the Zeeman term polarizes the system, we obtain
\begin{align}
\frac{\Delta v}{v}\sim A'''\left(1-D'''\frac{\chi}{T}\right)M^2 ,
\label{high}
\end{align}
where $M\sim \tanh[h/T]$ and $\chi=\partial M/\partial h$ are the magnetization and the magnetic susceptibilities, respectively, $h$ is magnetic field while $A'''$ and $D'''$ are constants. These asymptotic forms are applicable to general systems, beyond classical spin ice, and we can apply them to Yb$_{2}$Ti$_{2}$O$_{7}$. In Fig.~\ref{fit}(a), we show the fits to the asymptotic regions where Eqs.~(\ref{low}) and (\ref{high}) are valid for the data measured at $T=290$ mK shown in Fig. \ref{YbTOB dependcT}(a). The minimum in $\Delta v/v$, corresponds to the crossover between the paramagnetic and polarized phase. In Fig.~\ref{fit}(b), we attempt to describe the data at $T=20$ mK using Eq.~(\ref{high}). Clearly, at $20$ mK the response of the system is akin to that of the high-field behavior even for very small fields ($H\ge 0.1$ T). 

Note that, as seen from Eqs. (\ref{low}) and (\ref{high}), the sound-velocity change is typically given by the magnetization and magnetic susceptibility.\cite{PhysRevB.78.094406} The constant value of the sound velocity above 2.2 T (not shown) may indicate a fully polarized phase where magnetization is saturated. This is consistent with other experiments exhibiting well developed spin waves about a fully polarized state in the presence of such magnetic fields. \cite{ross2011quantum}  However, our central observation is the fact that this saturation is almost complete at around $0.1$ T ($0.2$ T) for 20 mK (120 mK) as is evident from Fig. \ref{YbTOB dependcT}. Based on this observation, we can infer that at low temperatures ($T< 170$ mK) the system is highly susceptible to a ferromagnetic phase with finite magnetic polarization. At higher temperature, the system enters a paramagnetic phase and requires finite fields to polarize. This critical field strength correlates with the positions of the minima observed in $\Delta v/v$ marked in Fig. \ref{YbTOB dependcT}. 

We can try to interpolate between the two asymptotic descriptions [Eqs.~(\ref{low}) and (\ref{high})] within a Curie-Weiss mean-field theory. For that, we used Eq.~(\ref{high}), but with the magnetization being determined self-consistently,
\begin{align}
M=\tanh\left[(x M+ g h)/T\right].
\label{eq_mag}
\end{align} 
Such an interpolation gives a good fit as shown in Fig.~\ref{fit}(c) for $T=290$ mK. Further, this mean-field theory predicts that the field required for reaching the polarized phase, characterized by the position of the minima (denoted by arrows in Figs.~\ref{YbTOB dependcT} and \ref{YbTOB dependcL}), must be proportional to temperature. Indeed, the position of the minima in $\Delta v/v$ grows approximately linear away from the zero-field transition (inset of Fig.~\ref{YbTOB dependcL}).

\section{Conclusion}

In this work, we have exploited the physics of magneto-elastic coupling to probe the magnetic behavior in the candidate quantum-spin-ice material Yb$_2$Ti$_2$O$_7$ using ultrasonic elastic waves by measuring the sound velocity and attenuation. Our results show the development of spin correlations with decreasing temperatures and indicate the presence of a cooperative paramagnetic phase that extends down about 0.17 K. Upon further cooling, both sound velocity and attenuation show non-monotonic features with a concomitant clear anomaly in the specific heat. The thermal expansion exhibits as well an anomaly at this temperature. The sound velocity and attenuation also show thermal hysteresis below this temperature. Measurements of the acoustic properties in magnetic fields indicate that the system is susceptible to ferromagnetic order for small magnetic fields above 0.17  K. Below this temperature, the system is best interpreted as being in a ferromagnetic phase. Hence, we suggest that Yb$_2$Ti$_2$O$_7$ evolves from a cooperative paramagnet to a ferromagnet around 0.17 K through a first order phase transition.

However, since the sound velocity can only probe  ${\bf q}$  = 0 correlations the actual nature of the magnetic order may differ from that of a ${\bf q}$ = 0 ferromagnet. Further, in the light of the recent proposal of a possible Coulomb ferromagnetic phase for this compound,\cite{savary2012coulombic} which, along with the magnetic order, supports an emergent photon, we note that, unfortunately, the present probe is unlikely to distinguish between a regular and a Coulomb ferromagnet. More microscopic probes are needed to address this issue. It might offer an interesting avenue of future research.

Finally, it may be of interest to point out a possible effect in ultrasonic measurements near a first order phase transition from a paramagnet to a ferromagnet. The ferromagnet so obtained will have domain walls. The elastic wave modulates the local exchange couplings, $J_{ij}$, for the spins.\cite{PhysRevB.83.184421} The sound waves enhance the fluctuation of the domain walls by locally altering the magnetic exchange energy of the bonds on which the domain walls lie. This then would directly affect the dynamics in the ferromagnetic phase and the thermal hysteresis near the transition. Therefore, it would be of interest to see if such frequency-dependent effects can be observed in ultrasound measurements.

\begin{acknowledgments}
We thank K. A. Ross for useful discussion. We acknowledge the support of the HLD at HZDR, member of the European Magnetic Field Laboratory (EMFL). The research has been supported by the DFG through SFB 1143 and WU 595/3-1 (Emmy Noether programme). We thank C.G.F. Blum for x-ray diffraction,  Laue back scattering and orientation of the samples. A.A.Z. acknowledges the support from the Institute for Chemistry of V.~N.~Karazin Kharkov National University.
\end{acknowledgments}
\appendix
\section{Specific heat and energy of classical spin ice within single-tetrahedron approximation}
\label{appen_spen}

For the classical spin ice, the energy per spin, $s$, and the specific heat are given by:\cite{PhysRevB.83.184421}
\begin{align}
\varepsilon_{spin}&=\frac{1}{N_s}\langle \sum_{\langle ij\rangle}  s^\alpha_iJ^{\alpha\beta}s^\beta_j\rangle\nonumber\\
c_{spin}&=\frac{1}{N_sT^2}\left<\left<\left(\sum_{\langle ij\rangle}s^\alpha_iJ^{\alpha\beta}s^\beta_j\right)^2\right>\right>,
\label{eq_cspin}
\end{align}
where ${N_s}$ is the total number of spins. For quantum spin ice there are additional transverse terms. The generalization of above expressions is straightforward in that case.

We can estimate the specific heat and energy for the classical spin-ice model within the single-tetrahedron approximation.\cite{timonin2011spin,zvyagin2013new,PhysRevB.84.144435} 

The partition function is given by:
\begin{align}
Z\propto \left[e^{2J/T}\left(6+8e^{-2J/T}+2e^{-8J/T}\right)\right]^{N_s/2},
\end{align} 
which gives
\begin{align}
\varepsilon_{spin} =-\frac{1}{N_s}\frac{\partial\ln Z}{\partial\beta}=-J+\frac{4J\left(e^{-2J/T}+e^{-8J/T}\right)}{\left(3+4e^{-2J/T}+e^{-8J/T}\right)}
\label{eq_en}
\end{align}
and
\begin{equation}
c_{spin}=\frac{\partial\varepsilon_{spin}}{\partial T}={24J^2e^{-2J/T}\left(1-2e^{-2J/T} +3e^{-4J/T}\right)\over T^2\left(3+e^{-2J/T} -e^{-4J/T}+e^{-6J/T}\right)^2}, \ 
\label{eq_cv} 
\end{equation}
where $J$ is the effective exchange constant and we use units with $k_B=1$.  The internal energy can be easily calculated. 

\section{Derivation of the sound-velocity change in presence of magnetic field}
\label{appen_derive}
For Eq. (\ref{eq1}), we obtain
\begin{equation}
{\Delta v\over v} \approx A (\varepsilon_{spin} -DTc_{spin}) \ , 
\end{equation}
where $\varepsilon_{spin}$ and  $c_{spin}$ are given by Eq.~(\ref{eq_cspin}) and for any operator $\mathcal{O}$,
\begin{align}
\langle\mathcal{O}\rangle&=\frac{1}{Z}Tr\left[e^{-\beta\mathcal{H}}\mathcal{O}\right]\nonumber\\
\langle\langle\mathcal{O}^2\rangle\rangle&=\frac{1}{Z}Tr\left[e^{-\beta\mathcal{H}}\mathcal{O}^2\right]-\langle\mathcal{O}\rangle^2 ,
\end{align}
with 
\begin{align}
\mathcal{H}=\sum_{\langle ij\rangle}J^{\alpha\beta}s^\alpha_is^\beta_j-\sum_i{\bf h}\cdot{\bf s}_i .
\end{align}
For low fields and low temperatures $(\beta h<1, \beta j>1)$, we obtain, by expanding the Zeeman term in the Hamiltonian,
\begin{align}
\varepsilon_{spin}&=\left.\varepsilon_{spin}\right|_{\bf h=0}+\frac{\beta^2h^2}{2N_s}\left<\left(\sum_{\langle ij\rangle} J^{\alpha\beta}s^\alpha_is^\beta_j\right)\left(\sum_i{\bf \hat h\cdot s}_i\right)^2\right>\nonumber\\
&\sim\left.\varepsilon_{spin}\right|_{\bf h=0} + \chi^2h^2
\end{align}
and
\begin{align}
c_{spin}&=\left.c_{spin}\right|_{\bf h=0}+\frac{\beta^2h^2}{2N_s}\left<\left<\left(\sum_{\langle ij\rangle} J^{\alpha\beta}s^\alpha_is^\beta_j\right)^2\left(\sum_i{\bf \hat h\cdot s}_i\right)^2\right>\right>\nonumber\\
&\sim\left.c_{spin}\right|_{\bf h=0} + \frac{\chi^4h^2}{T^2}.
\end{align}
Adding the two contributions, we obtain,
\begin{align}
\left.\frac{\Delta v}{v}\right|_{\bf h}=\left.\frac{\Delta v}{v}\right|_{\bf h=0}+A'\left(\chi^2-D'\frac{\chi^4}{T}\right)^2 .
\end{align}
Thus, the magnetic-field dependence in this regime is of the form given by Eq.~(\ref{low}).

For high fields $(\beta h>1, h/J>1)$, the system is essentially in a polarized phase and in leading order of calculating the expectation values of spin correlations we can neglect the spin Hamiltonian to obtain
\begin{align}
\varepsilon_{spin}\sim M^2;~~~~~~ c_{spin}\sim \frac{M^2\chi}{T^2} ,
\end{align}
where, $M\sim \tanh[h/T]$ is the high-field magnetization and $\chi=\partial M/\partial h$ is the corresponding magnetic susceptibility. This gives
\begin{align}
\left.\frac{\Delta v}{v}\right|_{\bf h}\sim A'''\left(1-D'''\frac{\chi}{T}\right)M^2 ,
\end{align}
which is given in Eq.~(\ref{high}).
\section{Fitting parameters used in Figure \ref{fit}}
\label{appen_fit}

Here we provide the fitting parameters (up to arbitrary shifts) for the different panels of Fig. 5.

\paragraph{Figure 5(a).} For the blue curve at low magnetic fields, the fitting parameter [with reference to Eq.~(\ref{low})] is $A''\approx 4.2\times 10^{-2}$. For the red curve at the near polarized phase, the fitting parameters [with reference to Eq.~(\ref{high})] are $A'''\approx 1.0 \times 10^{-4}$ and $D'''\approx 4.5\times 10^{-2}$.

\textcolor{black}{\paragraph{ Figure 5(b).} For the red curve, the fitting parameters are $A'''\approx8\times 10^{-5}$ and $D'''\approx3.6\times 10^{-2}$.}

\textcolor{black}{\paragraph{Figure 5(c).} The fitting parameters for the red curve [with reference to Eq.~(\ref{high}) where the self consistent magnetization is obtained from Eq.~(\ref{eq_mag})] are given by $A'''\approx0.3 \times 10^{-4}, D'''\approx1.6, x\approx0.1$, and $g\approx2.5$.  }

\textcolor{black}{Note that near the polarized phase the second term in Eq.~(\ref{high}) is subdominant for the panels (a) and (b), as expected. This is evident for the small values of $D'''$ for these two figures. For panel (c), $D'''$ is a really different constant, although the expression has the same form as of Eq.~(\ref{high}) which is why we have used the same notation for the coupling constants.}


\bibliography{reference}

\begin{thebibliography}{46}%
\makeatletter
\providecommand \@ifxundefined [1]{%
 \@ifx{#1\undefined}
}%
\providecommand \@ifnum [1]{%
 \ifnum #1\expandafter \@firstoftwo
 \else \expandafter \@secondoftwo
 \fi
}%
\providecommand \@ifx [1]{%
 \ifx #1\expandafter \@firstoftwo
 \else \expandafter \@secondoftwo
 \fi
}%
\providecommand \natexlab [1]{#1}%
\providecommand \enquote  [1]{``#1''}%
\providecommand \bibnamefont  [1]{#1}%
\providecommand \bibfnamefont [1]{#1}%
\providecommand \citenamefont [1]{#1}%
\providecommand \href@noop [0]{\@secondoftwo}%
\providecommand \href [0]{\begingroup \@sanitize@url \@href}%
\providecommand \@href[1]{\@@startlink{#1}\@@href}%
\providecommand \@@href[1]{\endgroup#1\@@endlink}%
\providecommand \@sanitize@url [0]{\catcode `\\12\catcode `\$12\catcode
  `\&12\catcode `\#12\catcode `\^12\catcode `\_12\catcode `\%12\relax}%
\providecommand \@@startlink[1]{}%
\providecommand \@@endlink[0]{}%
\providecommand \url  [0]{\begingroup\@sanitize@url \@url }%
\providecommand \@url [1]{\endgroup\@href {#1}{\urlprefix }}%
\providecommand \urlprefix  [0]{URL }%
\providecommand \Eprint [0]{\href }%
\@ifxundefined \urlstyle {%
  \providecommand \doi  [0]{\begingroup \@sanitize@url \@doi}%
  \providecommand \@doi [1]{\endgroup \@@startlink {\doibase
  #1}doi:\discretionary {}{}{}#1\@@endlink }%
}{%
  \providecommand \doi  [0]{doi:\discretionary{}{}{}\begingroup
  \urlstyle{rm}\Url }%
}%
\providecommand \doibase [0]{http://dx.doi.org/}%
\providecommand \Doi [0]{\begingroup \@sanitize@url \@Doi }%
\providecommand \@Doi  [1]{\endgroup\@@startlink{\doibase#1}\@@Doi}%
\providecommand \@@Doi [1]{#1\@@endlink}%
\providecommand \selectlanguage [0]{\@gobble}%
\providecommand \bibinfo  [0]{\@secondoftwo}%
\providecommand \bibfield  [0]{\@secondoftwo}%
\providecommand \translation [1]{[#1]}%
\providecommand \BibitemOpen [0]{}%
\providecommand \bibitemStop [0]{}%
\providecommand \bibitemNoStop [0]{.\EOS\space}%
\providecommand \EOS [0]{\spacefactor3000\relax}%
\providecommand \BibitemShut  [1]{\csname bibitem#1\endcsname}%
\bibitem [{\citenamefont {Hermele}\ \emph {et~al.}(2004)\citenamefont
  {Hermele}, \citenamefont {Fisher},\ and\ \citenamefont
  {Balents}}]{hermele2004pyrochlore}%
  \BibitemOpen
  \bibfield  {author} {\bibinfo {author} {\bibfnamefont {M.}~\bibnamefont
  {Hermele}}, \bibinfo {author} {\bibfnamefont {M.~P.~A.}\ \bibnamefont
  {Fisher}}, \ and\ \bibinfo {author} {\bibfnamefont {L.}~\bibnamefont
  {Balents}},\ }\href
  {http://journals.aps.org/prb/abstract/10.1103/PhysRevB.69.064404} {\bibfield
  {journal} {\bibinfo  {journal} {Phys. Rev. B},\ }\textbf {\bibinfo {volume}
  {69}},\ \bibinfo {pages} {064404} (\bibinfo {year} {2004})}\BibitemShut
  {NoStop}%
\bibitem [{\citenamefont {Moessner}\ and\ \citenamefont
  {Sondhi}(2003)}]{PhysRevB.68.054405}%
  \BibitemOpen
  \bibfield  {author} {\bibinfo {author} {\bibfnamefont {R.}~\bibnamefont
  {Moessner}}\ and\ \bibinfo {author} {\bibfnamefont {S.~L.}\ \bibnamefont
  {Sondhi}},\ }\href {http://link.aps.org/doi/10.1103/PhysRevB.68.054405}
  {\bibfield  {journal} {\bibinfo  {journal} {Phys. Rev. B},\ }\textbf
  {\bibinfo {volume} {68}},\ \bibinfo {pages} {054405} (\bibinfo {year}
  {2003})}\BibitemShut {NoStop}%
\bibitem [{\citenamefont {Gingras}\ and\ \citenamefont
  {McClarty}(2014)}]{gingras2014quantum}%
  \BibitemOpen
  \bibfield  {author} {\bibinfo {author} {\bibfnamefont {M.~J.}\ \bibnamefont
  {Gingras}}\ and\ \bibinfo {author} {\bibfnamefont {P.~A.}\ \bibnamefont
  {McClarty}},\ }\href {http://iopscience.iop.org/0034-4885/77/5/056501}
  {\bibfield  {journal} {\bibinfo  {journal} {Rep. Prog. Phys.},\ }\textbf
  {\bibinfo {volume} {77}},\ \bibinfo {pages} {056501} (\bibinfo {year}
  {2014})}\BibitemShut {NoStop}%
\bibitem [{\citenamefont {Savary}\ and\ \citenamefont
  {Balents}(2012)}]{savary2012coulombic}%
  \BibitemOpen
  \bibfield  {author} {\bibinfo {author} {\bibfnamefont {L.}~\bibnamefont
  {Savary}}\ and\ \bibinfo {author} {\bibfnamefont {L.}~\bibnamefont
  {Balents}},\ }\href
  {http://journals.aps.org/prl/abstract/10.1103/PhysRevLett.108.037202}
  {\bibfield  {journal} {\bibinfo  {journal} {Phys. Rev. Lett.},\ }\textbf
  {\bibinfo {volume} {108}},\ \bibinfo {pages} {037202} (\bibinfo {year}
  {2012})}\BibitemShut {NoStop}%
\bibitem [{\citenamefont {Castelnovo}\ \emph {et~al.}(2008)\citenamefont
  {Castelnovo}, \citenamefont {Moessner},\ and\ \citenamefont
  {Sondhi}}]{castelnovo2008magnetic}%
  \BibitemOpen
  \bibfield  {author} {\bibinfo {author} {\bibfnamefont {C.}~\bibnamefont
  {Castelnovo}}, \bibinfo {author} {\bibfnamefont {R.}~\bibnamefont
  {Moessner}}, \ and\ \bibinfo {author} {\bibfnamefont {S.~L.}\ \bibnamefont
  {Sondhi}},\ }\href
  {http://www.nature.com/nature/journal/v451/n7174/abs/nature06433.html}
  {\bibfield  {journal} {\bibinfo  {journal} {Nature},\ }\textbf {\bibinfo
  {volume} {451}},\ \bibinfo {pages} {42} (\bibinfo {year} {2008})}\BibitemShut
  {NoStop}%
\bibitem [{\citenamefont {Shannon}\ \emph {et~al.}(2012)\citenamefont
  {Shannon}, \citenamefont {Sikora}, \citenamefont {Pollmann}, \citenamefont
  {Penc},\ and\ \citenamefont {Fulde}}]{PhysRevLett.108.067204}%
  \BibitemOpen
  \bibfield  {author} {\bibinfo {author} {\bibfnamefont {N.}~\bibnamefont
  {Shannon}}, \bibinfo {author} {\bibfnamefont {O.}~\bibnamefont {Sikora}},
  \bibinfo {author} {\bibfnamefont {F.}~\bibnamefont {Pollmann}}, \bibinfo
  {author} {\bibfnamefont {K.}~\bibnamefont {Penc}}, \ and\ \bibinfo {author}
  {\bibfnamefont {P.}~\bibnamefont {Fulde}},\ }\href
  {http://link.aps.org/doi/10.1103/PhysRevLett.108.067204} {\bibfield
  {journal} {\bibinfo  {journal} {Phys. Rev. Lett.},\ }\textbf {\bibinfo
  {volume} {108}},\ \bibinfo {pages} {067204} (\bibinfo {year}
  {2012})}\BibitemShut {NoStop}%
\bibitem [{\citenamefont {Benton}\ \emph {et~al.}(2012)\citenamefont {Benton},
  \citenamefont {Sikora},\ and\ \citenamefont {Shannon}}]{PhysRevB.86.075154}%
  \BibitemOpen
  \bibfield  {author} {\bibinfo {author} {\bibfnamefont {O.}~\bibnamefont
  {Benton}}, \bibinfo {author} {\bibfnamefont {O.}~\bibnamefont {Sikora}}, \
  and\ \bibinfo {author} {\bibfnamefont {N.}~\bibnamefont {Shannon}},\ }\href
  {http://link.aps.org/doi/10.1103/PhysRevB.86.075154} {\bibfield  {journal}
  {\bibinfo  {journal} {Phys. Rev. B},\ }\textbf {\bibinfo {volume} {86}},\
  \bibinfo {pages} {075154} (\bibinfo {year} {2012})}\BibitemShut {NoStop}%
\bibitem [{\citenamefont {Huse}\ \emph {et~al.}(2003)\citenamefont {Huse},
  \citenamefont {Krauth}, \citenamefont {Moessner},\ and\ \citenamefont
  {Sondhi}}]{PhysRevLett.91.167004}%
  \BibitemOpen
  \bibfield  {author} {\bibinfo {author} {\bibfnamefont {D.~A.}\ \bibnamefont
  {Huse}}, \bibinfo {author} {\bibfnamefont {W.}~\bibnamefont {Krauth}},
  \bibinfo {author} {\bibfnamefont {R.}~\bibnamefont {Moessner}}, \ and\
  \bibinfo {author} {\bibfnamefont {S.~L.}\ \bibnamefont {Sondhi}},\ }\href
  {http://link.aps.org/doi/10.1103/PhysRevLett.91.167004} {\bibfield  {journal}
  {\bibinfo  {journal} {Phys. Rev. Lett.},\ }\textbf {\bibinfo {volume} {91}},\
  \bibinfo {pages} {167004} (\bibinfo {year} {2003})}\BibitemShut {NoStop}%
\bibitem [{\citenamefont {Henley}(2010)}]{henley2010coulomb}%
  \BibitemOpen
  \bibfield  {author} {\bibinfo {author} {\bibfnamefont {C.~L.}\ \bibnamefont
  {Henley}},\ }\href
  {http://www.annualreviews.org/doi/abs/10.1146/annurev-conmatphys-070909-104138}
  {\bibfield  {journal} {\bibinfo  {journal} {Annu. Rev. Condens. Matter
  Phys.},\ }\textbf {\bibinfo {volume} {1}},\ \bibinfo {pages} {179} (\bibinfo
  {year} {2010})}\BibitemShut {NoStop}%
\bibitem [{\citenamefont {Chang}\ \emph {et~al.}(2012)\citenamefont {Chang},
  \citenamefont {Onoda}, \citenamefont {Su}, \citenamefont {Kao}, \citenamefont
  {Tsuei}, \citenamefont {Yasui}, \citenamefont {Kakurai},\ and\ \citenamefont
  {Lees}}]{chang2012higgs}%
  \BibitemOpen
  \bibfield  {author} {\bibinfo {author} {\bibfnamefont {L.-J.}\ \bibnamefont
  {Chang}}, \bibinfo {author} {\bibfnamefont {S.}~\bibnamefont {Onoda}},
  \bibinfo {author} {\bibfnamefont {Y.}~\bibnamefont {Su}}, \bibinfo {author}
  {\bibfnamefont {Y.-J.}\ \bibnamefont {Kao}}, \bibinfo {author} {\bibfnamefont
  {K.-D.}\ \bibnamefont {Tsuei}}, \bibinfo {author} {\bibfnamefont
  {Y.}~\bibnamefont {Yasui}}, \bibinfo {author} {\bibfnamefont
  {K.}~\bibnamefont {Kakurai}}, \ and\ \bibinfo {author} {\bibfnamefont
  {M.~R.}\ \bibnamefont {Lees}},\ }\href
  {http://www.nature.com/ncomms/journal/v3/n8/abs/ncomms1989.html} {\bibfield
  {journal} {\bibinfo  {journal} {Nat. Commun.},\ }\textbf {\bibinfo {volume}
  {3}},\ \bibinfo {pages} {992} (\bibinfo {year} {2012})}\BibitemShut {NoStop}%
\bibitem [{\citenamefont {Applegate}\ \emph {et~al.}(2012)\citenamefont
  {Applegate}, \citenamefont {Hayre}, \citenamefont {Singh}, \citenamefont
  {Lin}, \citenamefont {Day},\ and\ \citenamefont
  {Gingras}}]{applegate2012vindication}%
  \BibitemOpen
  \bibfield  {author} {\bibinfo {author} {\bibfnamefont {R.}~\bibnamefont
  {Applegate}}, \bibinfo {author} {\bibfnamefont {N.~R.}\ \bibnamefont
  {Hayre}}, \bibinfo {author} {\bibfnamefont {R.~R.~P.}\ \bibnamefont {Singh}},
  \bibinfo {author} {\bibfnamefont {T.}~\bibnamefont {Lin}}, \bibinfo {author}
  {\bibfnamefont {A.~G.~R.}\ \bibnamefont {Day}}, \ and\ \bibinfo {author}
  {\bibfnamefont {M.~J.~P.}\ \bibnamefont {Gingras}},\ }\href
  {http://journals.aps.org/prl/abstract/10.1103/PhysRevLett.109.097205}
  {\bibfield  {journal} {\bibinfo  {journal} {Phys. Rev. Lett.},\ }\textbf
  {\bibinfo {volume} {109}},\ \bibinfo {pages} {097205} (\bibinfo {year}
  {2012})}\BibitemShut {NoStop}%
\bibitem [{\citenamefont {Thompson}\ \emph {et~al.}(2011)\citenamefont
  {Thompson}, \citenamefont {McClarty}, \citenamefont {R{\o}nnow},
  \citenamefont {Regnault}, \citenamefont {Sorge},\ and\ \citenamefont
  {Gingras}}]{thompson2011rods}%
  \BibitemOpen
  \bibfield  {author} {\bibinfo {author} {\bibfnamefont {J.~D.}\ \bibnamefont
  {Thompson}}, \bibinfo {author} {\bibfnamefont {P.~A.}\ \bibnamefont
  {McClarty}}, \bibinfo {author} {\bibfnamefont {H.~M.}\ \bibnamefont
  {R{\o}nnow}}, \bibinfo {author} {\bibfnamefont {L.~P.}\ \bibnamefont
  {Regnault}}, \bibinfo {author} {\bibfnamefont {A.}~\bibnamefont {Sorge}}, \
  and\ \bibinfo {author} {\bibfnamefont {M.~J.~P.}\ \bibnamefont {Gingras}},\
  }\href {http://journals.aps.org/prl/abstract/10.1103/PhysRevLett.106.187202}
  {\bibfield  {journal} {\bibinfo  {journal} {Phy. Rev. Lett.},\ }\textbf
  {\bibinfo {volume} {106}},\ \bibinfo {pages} {187202} (\bibinfo {year}
  {2011})}\BibitemShut {NoStop}%
\bibitem [{\citenamefont {Ross}\ \emph
  {et~al.}(2011){\natexlab{a}}\citenamefont {Ross}, \citenamefont {Savary},
  \citenamefont {Gaulin},\ and\ \citenamefont {Balents}}]{ross2011quantum}%
  \BibitemOpen
  \bibfield  {author} {\bibinfo {author} {\bibfnamefont {K.~A.}\ \bibnamefont
  {Ross}}, \bibinfo {author} {\bibfnamefont {L.}~\bibnamefont {Savary}},
  \bibinfo {author} {\bibfnamefont {B.~D.}\ \bibnamefont {Gaulin}}, \ and\
  \bibinfo {author} {\bibfnamefont {L.}~\bibnamefont {Balents}},\ }\href
  {http://journals.aps.org/prx/abstract/10.1103/PhysRevX.1.021002} {\bibfield
  {journal} {\bibinfo  {journal} {Phys. Rev. X},\ }\textbf {\bibinfo {volume}
  {1}},\ \bibinfo {pages} {021002} (\bibinfo {year}
  {2011}{\natexlab{a}})}\BibitemShut {NoStop}%
\bibitem [{\citenamefont {Yin}\ \emph {et~al.}(2013)\citenamefont {Yin},
  \citenamefont {Xia}, \citenamefont {Takano}, \citenamefont {Sullivan},
  \citenamefont {Li},\ and\ \citenamefont {Sun}}]{yin2013low}%
  \BibitemOpen
  \bibfield  {author} {\bibinfo {author} {\bibfnamefont {L.}~\bibnamefont
  {Yin}}, \bibinfo {author} {\bibfnamefont {J.~S.}\ \bibnamefont {Xia}},
  \bibinfo {author} {\bibfnamefont {Y.}~\bibnamefont {Takano}}, \bibinfo
  {author} {\bibfnamefont {N.~S.}\ \bibnamefont {Sullivan}}, \bibinfo {author}
  {\bibfnamefont {Q.~J.}\ \bibnamefont {Li}}, \ and\ \bibinfo {author}
  {\bibfnamefont {X.~F.}\ \bibnamefont {Sun}},\ }\href
  {http://journals.aps.org/prl/abstract/10.1103/PhysRevLett.110.137201}
  {\bibfield  {journal} {\bibinfo  {journal} {Phys. Rev. Lett.},\ }\textbf
  {\bibinfo {volume} {110}},\ \bibinfo {pages} {137201} (\bibinfo {year}
  {2013})}\BibitemShut {NoStop}%
\bibitem [{\citenamefont {Fennell}\ \emph {et~al.}(2012)\citenamefont
  {Fennell}, \citenamefont {Kenzelmann}, \citenamefont {Roessli}, \citenamefont
  {Haas},\ and\ \citenamefont {Cava}}]{fennell2012power}%
  \BibitemOpen
  \bibfield  {author} {\bibinfo {author} {\bibfnamefont {T.}~\bibnamefont
  {Fennell}}, \bibinfo {author} {\bibfnamefont {M.}~\bibnamefont {Kenzelmann}},
  \bibinfo {author} {\bibfnamefont {B.}~\bibnamefont {Roessli}}, \bibinfo
  {author} {\bibfnamefont {M.~K.}\ \bibnamefont {Haas}}, \ and\ \bibinfo
  {author} {\bibfnamefont {R.~J.}\ \bibnamefont {Cava}},\ }\href
  {http://journals.aps.org/prl/abstract/10.1103/PhysRevLett.109.017201}
  {\bibfield  {journal} {\bibinfo  {journal} {Phys. Rev. Lett.},\ }\textbf
  {\bibinfo {volume} {109}},\ \bibinfo {pages} {017201} (\bibinfo {year}
  {2012})}\BibitemShut {NoStop}%
\bibitem [{\citenamefont {Legl}\ \emph {et~al.}(2012)\citenamefont {Legl},
  \citenamefont {Krey}, \citenamefont {Dunsiger}, \citenamefont {Dabkowska},
  \citenamefont {Rodriguez}, \citenamefont {Luke},\ and\ \citenamefont
  {Pfleiderer}}]{legl2012vibrating}%
  \BibitemOpen
  \bibfield  {author} {\bibinfo {author} {\bibfnamefont {S.}~\bibnamefont
  {Legl}}, \bibinfo {author} {\bibfnamefont {C.}~\bibnamefont {Krey}}, \bibinfo
  {author} {\bibfnamefont {S.~R.}\ \bibnamefont {Dunsiger}}, \bibinfo {author}
  {\bibfnamefont {H.~A.}\ \bibnamefont {Dabkowska}}, \bibinfo {author}
  {\bibfnamefont {J.~A.}\ \bibnamefont {Rodriguez}}, \bibinfo {author}
  {\bibfnamefont {G.~M.}\ \bibnamefont {Luke}}, \ and\ \bibinfo {author}
  {\bibfnamefont {C.}~\bibnamefont {Pfleiderer}},\ }\href
  {http://journals.aps.org/prl/abstract/10.1103/PhysRevLett.109.047201}
  {\bibfield  {journal} {\bibinfo  {journal} {Phys. Rev. Lett.},\ }\textbf
  {\bibinfo {volume} {109}},\ \bibinfo {pages} {047201} (\bibinfo {year}
  {2012})}\BibitemShut {NoStop}%
\bibitem [{\citenamefont {Lee}\ \emph {et~al.}(2012)\citenamefont {Lee},
  \citenamefont {Onoda},\ and\ \citenamefont {Balents}}]{lee2012generic}%
  \BibitemOpen
  \bibfield  {author} {\bibinfo {author} {\bibfnamefont {S.~B.}\ \bibnamefont
  {Lee}}, \bibinfo {author} {\bibfnamefont {S.}~\bibnamefont {Onoda}}, \ and\
  \bibinfo {author} {\bibfnamefont {L.}~\bibnamefont {Balents}},\ }\href
  {http://journals.aps.org/prb/abstract/10.1103/PhysRevB.86.104412} {\bibfield
  {journal} {\bibinfo  {journal} {Phys. Rev. B},\ }\textbf {\bibinfo {volume}
  {86}},\ \bibinfo {pages} {104412} (\bibinfo {year} {2012})}\BibitemShut
  {NoStop}%
\bibitem [{\citenamefont {Pan}\ \emph {et~al.}(2015)\citenamefont {Pan},
  \citenamefont {Laurita}, \citenamefont {Ross}, \citenamefont {Kermarrec},
  \citenamefont {Gaulin},\ and\ \citenamefont {Armitage}}]{pan2015measure}%
  \BibitemOpen
  \bibfield  {author} {\bibinfo {author} {\bibfnamefont {L.}~\bibnamefont
  {Pan}}, \bibinfo {author} {\bibfnamefont {N.}~\bibnamefont {Laurita}},
  \bibinfo {author} {\bibfnamefont {K.~A.}\ \bibnamefont {Ross}}, \bibinfo
  {author} {\bibfnamefont {E.}~\bibnamefont {Kermarrec}}, \bibinfo {author}
  {\bibfnamefont {B.~D.}\ \bibnamefont {Gaulin}}, \ and\ \bibinfo {author}
  {\bibfnamefont {N.~P.}\ \bibnamefont {Armitage}},\ }\href
  {http://arxiv.org/abs/1501.05638} {\bibfield  {journal} {\bibinfo  {journal}
  {arXiv:1501.05638}} (\bibinfo {year} {2015})}\BibitemShut {NoStop}%
\bibitem [{\citenamefont {Canals}\ and\ \citenamefont
  {Lacroix}(1998)}]{PhysRevLett.80.2933}%
  \BibitemOpen
  \bibfield  {author} {\bibinfo {author} {\bibfnamefont {B.}~\bibnamefont
  {Canals}}\ and\ \bibinfo {author} {\bibfnamefont {C.}~\bibnamefont
  {Lacroix}},\ }\href {http://link.aps.org/doi/10.1103/PhysRevLett.80.2933}
  {\bibfield  {journal} {\bibinfo  {journal} {Phys. Rev. Lett.},\ }\textbf
  {\bibinfo {volume} {80}},\ \bibinfo {pages} {2933} (\bibinfo {year}
  {1998})}\BibitemShut {NoStop}%
\bibitem [{\citenamefont {Wan}\ and\ \citenamefont
  {Tchernyshyov}(2012)}]{PhysRevLett.108.247210}%
  \BibitemOpen
  \bibfield  {author} {\bibinfo {author} {\bibfnamefont {Y.}~\bibnamefont
  {Wan}}\ and\ \bibinfo {author} {\bibfnamefont {O.}~\bibnamefont
  {Tchernyshyov}},\ }\href
  {http://link.aps.org/doi/10.1103/PhysRevLett.108.247210} {\bibfield
  {journal} {\bibinfo  {journal} {Phys. Rev. Lett.},\ }\textbf {\bibinfo
  {volume} {108}},\ \bibinfo {pages} {247210} (\bibinfo {year}
  {2012})}\BibitemShut {NoStop}%
\bibitem [{\citenamefont {Hodges}\ \emph {et~al.}(2001)\citenamefont {Hodges},
  \citenamefont {Bonville}, \citenamefont {Forget}, \citenamefont {Rams},
  \citenamefont {Kr{\'o}las},\ and\ \citenamefont
  {Dhalenne}}]{hodges2001crystal}%
  \BibitemOpen
  \bibfield  {author} {\bibinfo {author} {\bibfnamefont {J.}~\bibnamefont
  {Hodges}}, \bibinfo {author} {\bibfnamefont {P.}~\bibnamefont {Bonville}},
  \bibinfo {author} {\bibfnamefont {A.}~\bibnamefont {Forget}}, \bibinfo
  {author} {\bibfnamefont {M.}~\bibnamefont {Rams}}, \bibinfo {author}
  {\bibfnamefont {K.}~\bibnamefont {Kr{\'o}las}}, \ and\ \bibinfo {author}
  {\bibfnamefont {G.}~\bibnamefont {Dhalenne}},\ }\href
  {http://iopscience.iop.org/0953-8984/13/41/318} {\bibfield  {journal}
  {\bibinfo  {journal} {J. Phys. Condens. Matter},\ }\textbf {\bibinfo {volume}
  {13}},\ \bibinfo {pages} {9301} (\bibinfo {year} {2001})}\BibitemShut
  {NoStop}%
\bibitem [{\citenamefont {Yasui}\ \emph {et~al.}(2003)\citenamefont {Yasui},
  \citenamefont {Soda}, \citenamefont {Iikubo}, \citenamefont {Ito},
  \citenamefont {Sato}, \citenamefont {Hamaguchi}, \citenamefont {Matsushita},
  \citenamefont {Wada}, \citenamefont {Takeuchi}, \citenamefont {Aso} \emph
  {et~al.}}]{yasui2003ferromagnetic}%
  \BibitemOpen
  \bibfield  {author} {\bibinfo {author} {\bibfnamefont {Y.}~\bibnamefont
  {Yasui}}, \bibinfo {author} {\bibfnamefont {M.}~\bibnamefont {Soda}},
  \bibinfo {author} {\bibfnamefont {S.}~\bibnamefont {Iikubo}}, \bibinfo
  {author} {\bibfnamefont {M.}~\bibnamefont {Ito}}, \bibinfo {author}
  {\bibfnamefont {M.}~\bibnamefont {Sato}}, \bibinfo {author} {\bibfnamefont
  {N.}~\bibnamefont {Hamaguchi}}, \bibinfo {author} {\bibfnamefont
  {T.}~\bibnamefont {Matsushita}}, \bibinfo {author} {\bibfnamefont
  {N.}~\bibnamefont {Wada}}, \bibinfo {author} {\bibfnamefont {T.}~\bibnamefont
  {Takeuchi}}, \bibinfo {author} {\bibfnamefont {N.}~\bibnamefont {Aso}},
  \emph {et~al.},\ }\href
  {https://www.jstage.jst.go.jp/article/jpsj/72/11/72_11_3014/_article}
  {\bibfield  {journal} {\bibinfo  {journal} {J. Phys. Soc. Jpn.},\ }\textbf
  {\bibinfo {volume} {72}},\ \bibinfo {pages} {3014} (\bibinfo {year}
  {2003})}\BibitemShut {NoStop}%
\bibitem [{\citenamefont {Chang}\ \emph {et~al.}(2014)\citenamefont {Chang},
  \citenamefont {Lees}, \citenamefont {Watanabe}, \citenamefont {Hillier},
  \citenamefont {Yasui},\ and\ \citenamefont {Onoda}}]{chang2014static}%
  \BibitemOpen
  \bibfield  {author} {\bibinfo {author} {\bibfnamefont {L.-J.}\ \bibnamefont
  {Chang}}, \bibinfo {author} {\bibfnamefont {M.~R.}\ \bibnamefont {Lees}},
  \bibinfo {author} {\bibfnamefont {I.}~\bibnamefont {Watanabe}}, \bibinfo
  {author} {\bibfnamefont {A.~D.}\ \bibnamefont {Hillier}}, \bibinfo {author}
  {\bibfnamefont {Y.}~\bibnamefont {Yasui}}, \ and\ \bibinfo {author}
  {\bibfnamefont {S.}~\bibnamefont {Onoda}},\ }\href
  {http://journals.aps.org/prb/abstract/10.1103/PhysRevB.89.184416} {\bibfield
  {journal} {\bibinfo  {journal} {Phys. Rev. B},\ }\textbf {\bibinfo {volume}
  {89}},\ \bibinfo {pages} {184416} (\bibinfo {year} {2014})}\BibitemShut
  {NoStop}%
\bibitem [{\citenamefont {Lhotel}\ \emph {et~al.}(2014)\citenamefont {Lhotel},
  \citenamefont {Giblin}, \citenamefont {Lees}, \citenamefont {Balakrishnan},
  \citenamefont {Chang},\ and\ \citenamefont {Yasui}}]{lhotel2014first}%
  \BibitemOpen
  \bibfield  {author} {\bibinfo {author} {\bibfnamefont {E.}~\bibnamefont
  {Lhotel}}, \bibinfo {author} {\bibfnamefont {S.~R.}\ \bibnamefont {Giblin}},
  \bibinfo {author} {\bibfnamefont {M.~R.}\ \bibnamefont {Lees}}, \bibinfo
  {author} {\bibfnamefont {G.}~\bibnamefont {Balakrishnan}}, \bibinfo {author}
  {\bibfnamefont {L.~J.}\ \bibnamefont {Chang}}, \ and\ \bibinfo {author}
  {\bibfnamefont {Y.}~\bibnamefont {Yasui}},\ }\href
  {http://link.aps.org/doi/10.1103/PhysRevB.89.224419} {\bibfield  {journal}
  {\bibinfo  {journal} {Phys. Rev. B},\ }\textbf {\bibinfo {volume} {89}},\
  \bibinfo {pages} {224419} (\bibinfo {year} {2014})}\BibitemShut {NoStop}%
\bibitem [{\citenamefont {Dun}\ \emph {et~al.}(2014)\citenamefont {Dun},
  \citenamefont {Lee}, \citenamefont {Choi}, \citenamefont {Hallas},
  \citenamefont {Wiebe}, \citenamefont {Gardner}, \citenamefont {Arrighi},
  \citenamefont {Freitas}, \citenamefont {Arevalo-Lopez}, \citenamefont
  {Attfield}, \citenamefont {Zhou},\ and\ \citenamefont
  {Cheng}}]{dun2014chemical}%
  \BibitemOpen
  \bibfield  {author} {\bibinfo {author} {\bibfnamefont {Z.~L.}\ \bibnamefont
  {Dun}}, \bibinfo {author} {\bibfnamefont {M.}~\bibnamefont {Lee}}, \bibinfo
  {author} {\bibfnamefont {E.~S.}\ \bibnamefont {Choi}}, \bibinfo {author}
  {\bibfnamefont {A.~M.}\ \bibnamefont {Hallas}}, \bibinfo {author}
  {\bibfnamefont {C.~R.}\ \bibnamefont {Wiebe}}, \bibinfo {author}
  {\bibfnamefont {J.~S.}\ \bibnamefont {Gardner}}, \bibinfo {author}
  {\bibfnamefont {E.}~\bibnamefont {Arrighi}}, \bibinfo {author} {\bibfnamefont
  {R.~S.}\ \bibnamefont {Freitas}}, \bibinfo {author} {\bibfnamefont {A.~M.}\
  \bibnamefont {Arevalo-Lopez}}, \bibinfo {author} {\bibfnamefont {J.~P.}\
  \bibnamefont {Attfield}}, \bibinfo {author} {\bibfnamefont {H.~D.}\
  \bibnamefont {Zhou}}, \ and\ \bibinfo {author} {\bibfnamefont {J.~G.}\
  \bibnamefont {Cheng}},\ }\href
  {http://link.aps.org/doi/10.1103/PhysRevB.89.064401} {\bibfield  {journal}
  {\bibinfo  {journal} {Phys. Rev. B},\ }\textbf {\bibinfo {volume} {89}},\
  \bibinfo {pages} {064401} (\bibinfo {year} {2014})}\BibitemShut {NoStop}%
\bibitem [{\citenamefont {Ross}\ \emph
  {et~al.}(2011){\natexlab{b}}\citenamefont {Ross}, \citenamefont
  {Yaraskavitch}, \citenamefont {Laver}, \citenamefont {Gardner}, \citenamefont
  {Quilliam}, \citenamefont {Meng}, \citenamefont {Kycia}, \citenamefont
  {Singh}, \citenamefont {Proffen}, \citenamefont {Dabkowska},\ and\
  \citenamefont {Gaulin}}]{PhysRevB.84.174442}%
  \BibitemOpen
  \bibfield  {author} {\bibinfo {author} {\bibfnamefont {K.~A.}\ \bibnamefont
  {Ross}}, \bibinfo {author} {\bibfnamefont {L.~R.}\ \bibnamefont
  {Yaraskavitch}}, \bibinfo {author} {\bibfnamefont {M.}~\bibnamefont {Laver}},
  \bibinfo {author} {\bibfnamefont {J.~S.}\ \bibnamefont {Gardner}}, \bibinfo
  {author} {\bibfnamefont {J.~A.}\ \bibnamefont {Quilliam}}, \bibinfo {author}
  {\bibfnamefont {S.}~\bibnamefont {Meng}}, \bibinfo {author} {\bibfnamefont
  {J.~B.}\ \bibnamefont {Kycia}}, \bibinfo {author} {\bibfnamefont {D.~K.}\
  \bibnamefont {Singh}}, \bibinfo {author} {\bibfnamefont {T.}~\bibnamefont
  {Proffen}}, \bibinfo {author} {\bibfnamefont {H.~A.}\ \bibnamefont
  {Dabkowska}}, \ and\ \bibinfo {author} {\bibfnamefont {B.~D.}\ \bibnamefont
  {Gaulin}},\ }\href {http://link.aps.org/doi/10.1103/PhysRevB.84.174442}
  {\bibfield  {journal} {\bibinfo  {journal} {Phys. Rev. B},\ }\textbf
  {\bibinfo {volume} {84}},\ \bibinfo {pages} {174442} (\bibinfo {year}
  {2011}{\natexlab{b}})}\BibitemShut {NoStop}%
\bibitem [{\citenamefont {Hodges}\ \emph {et~al.}(2002)\citenamefont {Hodges},
  \citenamefont {Bonville}, \citenamefont {Forget}, \citenamefont {Yaouanc},
  \citenamefont {Dalmas~de R\'eotier}, \citenamefont {Andr\'e}, \citenamefont
  {Rams}, \citenamefont {Kr\'olas}, \citenamefont {Ritter}, \citenamefont
  {Gubbens}, \citenamefont {Kaiser}, \citenamefont {King},\ and\ \citenamefont
  {Baines}}]{PhysRevLett.88.077204}%
  \BibitemOpen
  \bibfield  {author} {\bibinfo {author} {\bibfnamefont {J.~A.}\ \bibnamefont
  {Hodges}}, \bibinfo {author} {\bibfnamefont {P.}~\bibnamefont {Bonville}},
  \bibinfo {author} {\bibfnamefont {A.}~\bibnamefont {Forget}}, \bibinfo
  {author} {\bibfnamefont {A.}~\bibnamefont {Yaouanc}}, \bibinfo {author}
  {\bibfnamefont {P.}~\bibnamefont {Dalmas~de R\'eotier}}, \bibinfo {author}
  {\bibfnamefont {G.}~\bibnamefont {Andr\'e}}, \bibinfo {author} {\bibfnamefont
  {M.}~\bibnamefont {Rams}}, \bibinfo {author} {\bibfnamefont {K.}~\bibnamefont
  {Kr\'olas}}, \bibinfo {author} {\bibfnamefont {C.}~\bibnamefont {Ritter}},
  \bibinfo {author} {\bibfnamefont {P.~C.~M.}\ \bibnamefont {Gubbens}},
  \bibinfo {author} {\bibfnamefont {C.~T.}\ \bibnamefont {Kaiser}}, \bibinfo
  {author} {\bibfnamefont {P.~J.~C.}\ \bibnamefont {King}}, \ and\ \bibinfo
  {author} {\bibfnamefont {C.}~\bibnamefont {Baines}},\ }\href
  {http://link.aps.org/doi/10.1103/PhysRevLett.88.077204} {\bibfield  {journal}
  {\bibinfo  {journal} {Phys. Rev. Lett.},\ }\textbf {\bibinfo {volume} {88}},\
  \bibinfo {pages} {077204} (\bibinfo {year} {2002})}\BibitemShut {NoStop}%
\bibitem [{\citenamefont {Gardner}\ \emph {et~al.}(2004)\citenamefont
  {Gardner}, \citenamefont {Ehlers}, \citenamefont {Rosov}, \citenamefont
  {Erwin},\ and\ \citenamefont {Petrovic}}]{PhysRevB.70.180404}%
  \BibitemOpen
  \bibfield  {author} {\bibinfo {author} {\bibfnamefont {J.~S.}\ \bibnamefont
  {Gardner}}, \bibinfo {author} {\bibfnamefont {G.}~\bibnamefont {Ehlers}},
  \bibinfo {author} {\bibfnamefont {N.}~\bibnamefont {Rosov}}, \bibinfo
  {author} {\bibfnamefont {R.~W.}\ \bibnamefont {Erwin}}, \ and\ \bibinfo
  {author} {\bibfnamefont {C.}~\bibnamefont {Petrovic}},\ }\href
  {http://link.aps.org/doi/10.1103/PhysRevB.70.180404} {\bibfield  {journal}
  {\bibinfo  {journal} {Phys. Rev. B},\ }\textbf {\bibinfo {volume} {70}},\
  \bibinfo {pages} {180404} (\bibinfo {year} {2004})}\BibitemShut {NoStop}%
\bibitem [{\citenamefont {Ross}\ \emph {et~al.}(2009)\citenamefont {Ross},
  \citenamefont {Ruff}, \citenamefont {Adams}, \citenamefont {Gardner},
  \citenamefont {Dabkowska}, \citenamefont {Qiu}, \citenamefont {Copley},\ and\
  \citenamefont {Gaulin}}]{PhysRevLett.103.227202}%
  \BibitemOpen
  \bibfield  {author} {\bibinfo {author} {\bibfnamefont {K.~A.}\ \bibnamefont
  {Ross}}, \bibinfo {author} {\bibfnamefont {J.~P.~C.}\ \bibnamefont {Ruff}},
  \bibinfo {author} {\bibfnamefont {C.~P.}\ \bibnamefont {Adams}}, \bibinfo
  {author} {\bibfnamefont {J.~S.}\ \bibnamefont {Gardner}}, \bibinfo {author}
  {\bibfnamefont {H.~A.}\ \bibnamefont {Dabkowska}}, \bibinfo {author}
  {\bibfnamefont {Y.}~\bibnamefont {Qiu}}, \bibinfo {author} {\bibfnamefont
  {J.~R.~D.}\ \bibnamefont {Copley}}, \ and\ \bibinfo {author} {\bibfnamefont
  {B.~D.}\ \bibnamefont {Gaulin}},\ }\href
  {http://link.aps.org/doi/10.1103/PhysRevLett.103.227202} {\bibfield
  {journal} {\bibinfo  {journal} {Phys. Rev. Lett.},\ }\textbf {\bibinfo
  {volume} {103}},\ \bibinfo {pages} {227202} (\bibinfo {year}
  {2009})}\BibitemShut {NoStop}%
\bibitem [{\citenamefont {D'Ortenzio}\ \emph {et~al.}(2013)\citenamefont
  {D'Ortenzio}, \citenamefont {Dabkowska}, \citenamefont {Dunsiger},
  \citenamefont {Gaulin}, \citenamefont {Gingras}, \citenamefont {Goko},
  \citenamefont {Kycia}, \citenamefont {Liu}, \citenamefont {Medina},
  \citenamefont {Munsie}, \citenamefont {Pomaranski}, \citenamefont {Ross},
  \citenamefont {Uemura}, \citenamefont {Williams},\ and\ \citenamefont
  {Luke}}]{PhysRevB.88.134428}%
  \BibitemOpen
  \bibfield  {author} {\bibinfo {author} {\bibfnamefont {R.~M.}\ \bibnamefont
  {D'Ortenzio}}, \bibinfo {author} {\bibfnamefont {H.~A.}\ \bibnamefont
  {Dabkowska}}, \bibinfo {author} {\bibfnamefont {S.~R.}\ \bibnamefont
  {Dunsiger}}, \bibinfo {author} {\bibfnamefont {B.~D.}\ \bibnamefont
  {Gaulin}}, \bibinfo {author} {\bibfnamefont {M.~J.~P.}\ \bibnamefont
  {Gingras}}, \bibinfo {author} {\bibfnamefont {T.}~\bibnamefont {Goko}},
  \bibinfo {author} {\bibfnamefont {J.~B.}\ \bibnamefont {Kycia}}, \bibinfo
  {author} {\bibfnamefont {L.}~\bibnamefont {Liu}}, \bibinfo {author}
  {\bibfnamefont {T.}~\bibnamefont {Medina}}, \bibinfo {author} {\bibfnamefont
  {T.~J.}\ \bibnamefont {Munsie}}, \bibinfo {author} {\bibfnamefont
  {D.}~\bibnamefont {Pomaranski}}, \bibinfo {author} {\bibfnamefont {K.~A.}\
  \bibnamefont {Ross}}, \bibinfo {author} {\bibfnamefont {Y.~J.}\ \bibnamefont
  {Uemura}}, \bibinfo {author} {\bibfnamefont {T.~J.}\ \bibnamefont
  {Williams}}, \ and\ \bibinfo {author} {\bibfnamefont {G.~M.}\ \bibnamefont
  {Luke}},\ }\href {http://link.aps.org/doi/10.1103/PhysRevB.88.134428}
  {\bibfield  {journal} {\bibinfo  {journal} {Phys. Rev. B},\ }\textbf
  {\bibinfo {volume} {88}},\ \bibinfo {pages} {134428} (\bibinfo {year}
  {2013})}\BibitemShut {NoStop}%
\bibitem [{\citenamefont {Ross}\ \emph {et~al.}(2012)\citenamefont {Ross},
  \citenamefont {Proffen}, \citenamefont {Dabkowska}, \citenamefont {Quilliam},
  \citenamefont {Yaraskavitch}, \citenamefont {Kycia},\ and\ \citenamefont
  {Gaulin}}]{PhysRevB.86.174424}%
  \BibitemOpen
  \bibfield  {author} {\bibinfo {author} {\bibfnamefont {K.~A.}\ \bibnamefont
  {Ross}}, \bibinfo {author} {\bibfnamefont {T.}~\bibnamefont {Proffen}},
  \bibinfo {author} {\bibfnamefont {H.~A.}\ \bibnamefont {Dabkowska}}, \bibinfo
  {author} {\bibfnamefont {J.~A.}\ \bibnamefont {Quilliam}}, \bibinfo {author}
  {\bibfnamefont {L.~R.}\ \bibnamefont {Yaraskavitch}}, \bibinfo {author}
  {\bibfnamefont {J.~B.}\ \bibnamefont {Kycia}}, \ and\ \bibinfo {author}
  {\bibfnamefont {B.~D.}\ \bibnamefont {Gaulin}},\ }\href
  {http://link.aps.org/doi/10.1103/PhysRevB.86.174424} {\bibfield  {journal}
  {\bibinfo  {journal} {Phys. Rev. B},\ }\textbf {\bibinfo {volume} {86}},\
  \bibinfo {pages} {174424} (\bibinfo {year} {2012})}\BibitemShut {NoStop}%
\bibitem [{\citenamefont {Yaouanc}\ \emph {et~al.}(2011)\citenamefont
  {Yaouanc}, \citenamefont {Dalmas~de R\'eotier}, \citenamefont {Marin},\ and\
  \citenamefont {Glazkov}}]{PhysRevB.84.172408}%
  \BibitemOpen
  \bibfield  {author} {\bibinfo {author} {\bibfnamefont {A.}~\bibnamefont
  {Yaouanc}}, \bibinfo {author} {\bibfnamefont {P.}~\bibnamefont {Dalmas~de
  R\'eotier}}, \bibinfo {author} {\bibfnamefont {C.}~\bibnamefont {Marin}}, \
  and\ \bibinfo {author} {\bibfnamefont {V.}~\bibnamefont {Glazkov}},\ }\href
  {http://link.aps.org/doi/10.1103/PhysRevB.84.172408} {\bibfield  {journal}
  {\bibinfo  {journal} {Phys. Rev. B},\ }\textbf {\bibinfo {volume} {84}},\
  \bibinfo {pages} {172408} (\bibinfo {year} {2011})}\BibitemShut {NoStop}%
\bibitem [{\citenamefont {Fradkin}\ \emph {et~al.}(2004)\citenamefont
  {Fradkin}, \citenamefont {Huse}, \citenamefont {Moessner}, \citenamefont
  {Oganesyan},\ and\ \citenamefont {Sondhi}}]{PhysRevB.69.224415}%
  \BibitemOpen
  \bibfield  {author} {\bibinfo {author} {\bibfnamefont {E.}~\bibnamefont
  {Fradkin}}, \bibinfo {author} {\bibfnamefont {D.~A.}\ \bibnamefont {Huse}},
  \bibinfo {author} {\bibfnamefont {R.}~\bibnamefont {Moessner}}, \bibinfo
  {author} {\bibfnamefont {V.}~\bibnamefont {Oganesyan}}, \ and\ \bibinfo
  {author} {\bibfnamefont {S.~L.}\ \bibnamefont {Sondhi}},\ }\href
  {http://link.aps.org/doi/10.1103/PhysRevB.69.224415} {\bibfield  {journal}
  {\bibinfo  {journal} {Phys. Rev. B},\ }\textbf {\bibinfo {volume} {69}},\
  \bibinfo {pages} {224415} (\bibinfo {year} {2004})}\BibitemShut {NoStop}%
\bibitem [{\citenamefont {Erfanifam}\ \emph {et~al.}(2011)\citenamefont
  {Erfanifam}, \citenamefont {Zherlitsyn}, \citenamefont {Wosnitza},
  \citenamefont {Moessner}, \citenamefont {Petrenko}, \citenamefont
  {Balakrishnan},\ and\ \citenamefont {Zvyagin}}]{PhysRevB.84.220404}%
  \BibitemOpen
  \bibfield  {author} {\bibinfo {author} {\bibfnamefont {S.}~\bibnamefont
  {Erfanifam}}, \bibinfo {author} {\bibfnamefont {S.}~\bibnamefont
  {Zherlitsyn}}, \bibinfo {author} {\bibfnamefont {J.}~\bibnamefont
  {Wosnitza}}, \bibinfo {author} {\bibfnamefont {R.}~\bibnamefont {Moessner}},
  \bibinfo {author} {\bibfnamefont {O.~A.}\ \bibnamefont {Petrenko}}, \bibinfo
  {author} {\bibfnamefont {G.}~\bibnamefont {Balakrishnan}}, \ and\ \bibinfo
  {author} {\bibfnamefont {A.~A.}\ \bibnamefont {Zvyagin}},\ }\href
  {http://link.aps.org/doi/10.1103/PhysRevB.84.220404} {\bibfield  {journal}
  {\bibinfo  {journal} {Phys. Rev. B},\ }\textbf {\bibinfo {volume} {84}},\
  \bibinfo {pages} {220404} (\bibinfo {year} {2011})}\BibitemShut {NoStop}%
\bibitem [{\citenamefont {Erfanifam}\ \emph {et~al.}(2014)\citenamefont
  {Erfanifam}, \citenamefont {Zherlitsyn}, \citenamefont {Yasin}, \citenamefont
  {Skourski}, \citenamefont {Wosnitza}, \citenamefont {Zvyagin}, \citenamefont
  {McClarty}, \citenamefont {Moessner}, \citenamefont {Balakrishnan},\ and\
  \citenamefont {Petrenko}}]{PhysRevB.90.064409}%
  \BibitemOpen
  \bibfield  {author} {\bibinfo {author} {\bibfnamefont {S.}~\bibnamefont
  {Erfanifam}}, \bibinfo {author} {\bibfnamefont {S.}~\bibnamefont
  {Zherlitsyn}}, \bibinfo {author} {\bibfnamefont {S.}~\bibnamefont {Yasin}},
  \bibinfo {author} {\bibfnamefont {Y.}~\bibnamefont {Skourski}}, \bibinfo
  {author} {\bibfnamefont {J.}~\bibnamefont {Wosnitza}}, \bibinfo {author}
  {\bibfnamefont {A.~A.}\ \bibnamefont {Zvyagin}}, \bibinfo {author}
  {\bibfnamefont {P.}~\bibnamefont {McClarty}}, \bibinfo {author}
  {\bibfnamefont {R.}~\bibnamefont {Moessner}}, \bibinfo {author}
  {\bibfnamefont {G.}~\bibnamefont {Balakrishnan}}, \ and\ \bibinfo {author}
  {\bibfnamefont {O.~A.}\ \bibnamefont {Petrenko}},\ }\href
  {http://link.aps.org/doi/10.1103/PhysRevB.90.064409} {\bibfield  {journal}
  {\bibinfo  {journal} {Phys. Rev. B},\ }\textbf {\bibinfo {volume} {90}},\
  \bibinfo {pages} {064409} (\bibinfo {year} {2014})}\BibitemShut {NoStop}%
\bibitem [{\citenamefont {L{\"u}thi}(2007)}]{luthi2007physical}%
  \BibitemOpen
  \bibfield  {author} {\bibinfo {author} {\bibfnamefont {B.}~\bibnamefont
  {L{\"u}thi}},\ }\href
  {http://www.springer.com/de/book/9783540229100#otherversion=9783540721932}
  {\emph {\bibinfo {title} {Physical acoustics in the solid state}}},\ Vol.\
  \bibinfo {volume} {148}\ (\bibinfo  {publisher} {Springer Science \& Business
  Media},\ \bibinfo {year} {2007})\BibitemShut {NoStop}%
\bibitem [{\citenamefont {Rotter}\ \emph {et~al.}(1998)\citenamefont {Rotter},
  \citenamefont {M\"{u}ller}, \citenamefont {Gratz}, \citenamefont {Doerr},\
  and\ \citenamefont {Loewenhaupt}}]{rotter}%
  \BibitemOpen
  \bibfield  {author} {\bibinfo {author} {\bibfnamefont {M.}~\bibnamefont
  {Rotter}}, \bibinfo {author} {\bibfnamefont {H.}~\bibnamefont {M\"{u}ller}},
  \bibinfo {author} {\bibfnamefont {E.}~\bibnamefont {Gratz}}, \bibinfo
  {author} {\bibfnamefont {M.}~\bibnamefont {Doerr}}, \ and\ \bibinfo {author}
  {\bibfnamefont {M.}~\bibnamefont {Loewenhaupt}},\ }\href
  {http://scitation.aip.org/content/aip/journal/rsi/69/7/10.1063/1.1149009}
  {\bibfield  {journal} {\bibinfo  {journal} {Rev. Sci. Instrum.},\ }\textbf
  {\bibinfo {volume} {69}},\ \bibinfo {pages} {2742} (\bibinfo {year}
  {1998})}\BibitemShut {NoStop}%
\bibitem [{\citenamefont {Bhattacharjee}\ \emph {et~al.}(2011)\citenamefont
  {Bhattacharjee}, \citenamefont {Zherlitsyn}, \citenamefont {Chiatti},
  \citenamefont {Sytcheva}, \citenamefont {Wosnitza}, \citenamefont {Moessner},
  \citenamefont {Zhitomirsky}, \citenamefont {Lemmens}, \citenamefont
  {Tsurkan},\ and\ \citenamefont {Loidl}}]{PhysRevB.83.184421}%
  \BibitemOpen
  \bibfield  {author} {\bibinfo {author} {\bibfnamefont {S.}~\bibnamefont
  {Bhattacharjee}}, \bibinfo {author} {\bibfnamefont {S.}~\bibnamefont
  {Zherlitsyn}}, \bibinfo {author} {\bibfnamefont {O.}~\bibnamefont {Chiatti}},
  \bibinfo {author} {\bibfnamefont {A.}~\bibnamefont {Sytcheva}}, \bibinfo
  {author} {\bibfnamefont {J.}~\bibnamefont {Wosnitza}}, \bibinfo {author}
  {\bibfnamefont {R.}~\bibnamefont {Moessner}}, \bibinfo {author}
  {\bibfnamefont {M.~E.}\ \bibnamefont {Zhitomirsky}}, \bibinfo {author}
  {\bibfnamefont {P.}~\bibnamefont {Lemmens}}, \bibinfo {author} {\bibfnamefont
  {V.}~\bibnamefont {Tsurkan}}, \ and\ \bibinfo {author} {\bibfnamefont
  {A.}~\bibnamefont {Loidl}},\ }\href
  {http://link.aps.org/doi/10.1103/PhysRevB.83.184421} {\bibfield  {journal}
  {\bibinfo  {journal} {Phys. Rev. B},\ }\textbf {\bibinfo {volume} {83}},\
  \bibinfo {pages} {184421} (\bibinfo {year} {2011})}\BibitemShut {NoStop}%
\bibitem [{\citenamefont {Tachiki}\ and\ \citenamefont
  {Maekawa}(1974)}]{Tachiki01011974}%
  \BibitemOpen
  \bibfield  {author} {\bibinfo {author} {\bibfnamefont {M.}~\bibnamefont
  {Tachiki}}\ and\ \bibinfo {author} {\bibfnamefont {S.}~\bibnamefont
  {Maekawa}},\ }\href {http://ptp.oxfordjournals.org/content/51/1/1.abstract}
  {\bibfield  {journal} {\bibinfo  {journal} {Prog. Theor. Phys.},\ }\textbf
  {\bibinfo {volume} {51}},\ \bibinfo {pages} {1} (\bibinfo {year}
  {1974})}\BibitemShut {NoStop}%
\bibitem [{\citenamefont {Ryzhkin}(2005)}]{ryzhkin2005magnetic}%
  \BibitemOpen
  \bibfield  {author} {\bibinfo {author} {\bibfnamefont {I.}~\bibnamefont
  {Ryzhkin}},\ }\href {http://link.springer.com/article/10.1134/1.2103216}
  {\bibfield  {journal} {\bibinfo  {journal} {J. Exp. Theor. Phys.},\ }\textbf
  {\bibinfo {volume} {101}},\ \bibinfo {pages} {481} (\bibinfo {year}
  {2005})}\BibitemShut {NoStop}%
\bibitem [{\citenamefont {Bramwell}(2012)}]{bramwell2012generalized}%
  \BibitemOpen
  \bibfield  {author} {\bibinfo {author} {\bibfnamefont {S.~T.}\ \bibnamefont
  {Bramwell}},\ }\href
  {http://rsta.royalsocietypublishing.org/content/370/1981/5738.short}
  {\bibfield  {journal} {\bibinfo  {journal} {Phil. Trans. R. Soc. A},\
  }\textbf {\bibinfo {volume} {370}},\ \bibinfo {pages} {5738} (\bibinfo {year}
  {2012})}\BibitemShut {NoStop}%
\bibitem [{\citenamefont {Zvyagin}(2013)}]{zvyagin2013new}%
  \BibitemOpen
  \bibfield  {author} {\bibinfo {author} {\bibfnamefont {A.~A.}\ \bibnamefont
  {Zvyagin}},\ }\href
  {http://scitation.aip.org/content/aip/journal/ltp/39/11/10.1063/1.4826079}
  {\bibfield  {journal} {\bibinfo  {journal} {Low Temp. Phys.},\ }\textbf
  {\bibinfo {volume} {39}},\ \bibinfo {pages} {901} (\bibinfo {year}
  {2013})}\BibitemShut {NoStop}%
\bibitem [{\citenamefont {Yaouanc}\ \emph {et~al.}(2003)\citenamefont
  {Yaouanc}, \citenamefont {de~R{\'e}otier}, \citenamefont {Bonville},
  \citenamefont {Hodges}, \citenamefont {Gubbens}, \citenamefont {Kaiser},\
  and\ \citenamefont {Sakarya}}]{yaouanc2003spin}%
  \BibitemOpen
  \bibfield  {author} {\bibinfo {author} {\bibfnamefont {A.}~\bibnamefont
  {Yaouanc}}, \bibinfo {author} {\bibfnamefont {P.~D.}\ \bibnamefont
  {de~R{\'e}otier}}, \bibinfo {author} {\bibfnamefont {P.}~\bibnamefont
  {Bonville}}, \bibinfo {author} {\bibfnamefont {J.}~\bibnamefont {Hodges}},
  \bibinfo {author} {\bibfnamefont {P.}~\bibnamefont {Gubbens}}, \bibinfo
  {author} {\bibfnamefont {C.}~\bibnamefont {Kaiser}}, \ and\ \bibinfo {author}
  {\bibfnamefont {S.}~\bibnamefont {Sakarya}},\ }\href
  {http://www.sciencedirect.com/science/article/pii/S0921452602016642}
  {\bibfield  {journal} {\bibinfo  {journal} {Physica B},\ }\textbf {\bibinfo
  {volume} {326}},\ \bibinfo {pages} {456} (\bibinfo {year}
  {2003})}\BibitemShut {NoStop}%
\bibitem [{\citenamefont {Chiatti}\ \emph {et~al.}(2008)\citenamefont
  {Chiatti}, \citenamefont {Sytcheva}, \citenamefont {Wosnitza}, \citenamefont
  {Zherlitsyn}, \citenamefont {Zvyagin}, \citenamefont {Zapf}, \citenamefont
  {Jaime},\ and\ \citenamefont {Paduan-Filho}}]{PhysRevB.78.094406}%
  \BibitemOpen
  \bibfield  {author} {\bibinfo {author} {\bibfnamefont {O.}~\bibnamefont
  {Chiatti}}, \bibinfo {author} {\bibfnamefont {A.}~\bibnamefont {Sytcheva}},
  \bibinfo {author} {\bibfnamefont {J.}~\bibnamefont {Wosnitza}}, \bibinfo
  {author} {\bibfnamefont {S.}~\bibnamefont {Zherlitsyn}}, \bibinfo {author}
  {\bibfnamefont {A.~A.}\ \bibnamefont {Zvyagin}}, \bibinfo {author}
  {\bibfnamefont {V.~S.}\ \bibnamefont {Zapf}}, \bibinfo {author}
  {\bibfnamefont {M.}~\bibnamefont {Jaime}}, \ and\ \bibinfo {author}
  {\bibfnamefont {A.}~\bibnamefont {Paduan-Filho}},\ }\href
  {http://link.aps.org/doi/10.1103/PhysRevB.78.094406} {\bibfield  {journal}
  {\bibinfo  {journal} {Phys. Rev. B},\ }\textbf {\bibinfo {volume} {78}},\
  \bibinfo {pages} {094406} (\bibinfo {year} {2008})}\BibitemShut {NoStop}%
\bibitem [{\citenamefont {Timonin}(2011)}]{timonin2011spin}%
  \BibitemOpen
  \bibfield  {author} {\bibinfo {author} {\bibfnamefont {P.}~\bibnamefont
  {Timonin}},\ }\href
  {http://link.springer.com/article/10.1134/S1063776111080115} {\bibfield
  {journal} {\bibinfo  {journal} {J. Exp. Theor. Phys.},\ }\textbf {\bibinfo
  {volume} {113}},\ \bibinfo {pages} {251} (\bibinfo {year}
  {2011})}\BibitemShut {NoStop}%
\bibitem [{\citenamefont {Castelnovo}\ \emph {et~al.}(2011)\citenamefont
  {Castelnovo}, \citenamefont {Moessner},\ and\ \citenamefont
  {Sondhi}}]{PhysRevB.84.144435}%
  \BibitemOpen
  \bibfield  {author} {\bibinfo {author} {\bibfnamefont {C.}~\bibnamefont
  {Castelnovo}}, \bibinfo {author} {\bibfnamefont {R.}~\bibnamefont
  {Moessner}}, \ and\ \bibinfo {author} {\bibfnamefont {S.~L.}\ \bibnamefont
  {Sondhi}},\ }\href {http://link.aps.org/doi/10.1103/PhysRevB.84.144435}
  {\bibfield  {journal} {\bibinfo  {journal} {Phys. Rev. B},\ }\textbf
  {\bibinfo {volume} {84}},\ \bibinfo {pages} {144435} (\bibinfo {year}
  {2011})}\BibitemShut {NoStop}%
\end{thebibliography}%
\end{document}